\documentclass[reprint, superscriptaddress, longbibliography, bibnotes, amsmath,amssymb, aps, prb]{revtex4-2}
\usepackage[utf8]{inputenc}
\usepackage[english]{babel}
\usepackage[plainpages = false, pdfpagelabels, 
                 bookmarks,
                 bookmarksopen = true,
                 bookmarksnumbered = true,
                 linktocpage,
                 colorlinks = true,
                 linkcolor = blue,
                 urlcolor  = blue,
                 citecolor = blue,
                 anchorcolor = green,
                 hyperindex = true,
                 hyperfigures
                 ]{hyperref} 
\usepackage{amssymb}
\usepackage{amsmath}
	\DeclareMathOperator{\Span}{span}
	\newcommand{\spanof}[1]{\Span(\{#1\})}
\usepackage{enumerate}
\usepackage{physics}
\usepackage{bbold}
\usepackage{bm}
\usepackage{mathtools}
\usepackage{float}
\usepackage{graphicx}
	\graphicspath{{figures/}}
\usepackage{subcaption}
\usepackage{hyperref}

\addto\captionsenglish{

}
\AtBeginDocument{\pagenumbering{arabic}}

\begin{document}
\title{\texorpdfstring{Yu-Shiba-Rusinov States in Ising Superconductors}{}}
\author{Michael Hein}
\affiliation{Fachbereich Physik, University of Konstanz, D-78457 Konstanz, Germany}
\author{Juan Carlos Cuevas}
\affiliation{Departamento de Física Teórica de la Materia Condensada and Condensed Matter Physics Center (IFIMAC), Universidad Autónoma de Madrid, E-28049 Madrid, Spain}
\author{Wolfgang Belzig}
\affiliation{Fachbereich Physik, University of Konstanz, D-78457 Konstanz, Germany}

\begin{abstract}
The nature of the superconducting state in two-dimensional transition-metal dichalcogenides remains under active debate. A widely used description invokes so-called \textit{Ising superconductivity}. In this work, we investigate theoretically this pairing state by employing single magnetic impurities as local probes of the superconducting condensate. We analyze the formation of Yu-Shiba-Rusinov bound states in the presence of Ising spin-orbit coupling and an in-plane magnetic field to study how their spectral properties encode the underlying pairing structure. We identify distinct features in the bound-state spectrum and tunneling response that differentiate this system from conventional superconductors. Our results demonstrate that magnetic impurities provide a sensitive probe of the structure of the superconducting state and yield experimentally accessible signatures of unconventional aspects of Ising superconductivity.
\end{abstract}

\date{\today}
\maketitle

\section{Introduction}
\label{sec:Intro}
Two-dimensional transition-metal dichalcogenides (TMDs) provide a versatile platform for studying superconductivity in systems without inversion symmetry~\cite{ajayan_two-dimensional_2016}. In odd-layer compounds and in particular in monolayers, the crystal structure belongs to the point group $D_{3h}$ and preserves a horizontal mirror plane $\sigma_h$ while breaking inversion symmetry. As a consequence spin-orbit coupling generates an effective out-of-plane spin-splitting with opposite sign in the two valleys, commonly referred to as Ising spin-orbit coupling (ISOC)~\cite{Frigeri2004Sep,PhysRevB.93.180501,Xi2016Feb,Saito2016Feb}. In the normal state, this leads to spin-valley locking and valley-dependent Fermi surfaces at $K$ and $K'$ which have been experimentally resolved~\cite{Xiao2012May,Bawden2016May,Nakata2018May}.

When superconductivity emerges in these systems, ISOC strongly modifies the response to an in-plane magnetic field. In contrast to conventional BCS superconductors where Zeeman pair breaking suppresses superconductivity at the Pauli limit, Ising superconductors can remain robust far beyond this limit~\cite{doi:10.1126/science.aab2277,Saito2016Feb}. This unusual resilience has been widely interpreted as a consequence of spin-valley locking and the resulting protection of Cooper pairs against in-plane fields. Beyond this mechanism, a number of theoretical works have pointed out that Ising superconductors may host unconventional pairing correlations, including induced triplet components and field-dependent modifications of the excitation spectrum~\cite{PhysRevB.101.014510,Haim2020Dec,PhysRevB.106.184514}. In particular, the emergence of a magnetic-field-induced ``mirage'' gap at finite energies has been proposed as a characteristic spectroscopic signature of these systems~\cite{PhysRevLett.126.237001}.

Identifying clear experimental signatures that distinguish Ising superconductors from conventional superconductors remains an open challenge. A particularly sensitive probe of the superconducting state is provided by magnetic impurities. It is well established that a magnetic impurity embedded in a superconductor locally breaks Cooper pairs and induces Yu-Shiba-Rusinov (YSR) bound states inside the superconducting gap~\cite{Shiba1968Sep,PismaZhETF.9.146}. The properties of these subgap states depend sensitively on the spin structure and pairing symmetry of the host superconductor, making them a powerful local spectroscopic tool~\cite{RevModPhys.78.373,Morr2008}.

In the regime where the impurity hosts a well-defined local magnetic moment, the low-energy physics can be described by an effective exchange coupling between the impurity spin and the conduction electrons. Microscopically, this situation is described by the single-impurity Anderson model in the limit of large on-site Coulomb interaction where charge fluctuations are suppressed and a local moment forms. The validity of this description is controlled by the ratio between the Kondo temperature $T_{\rm K}$ and the superconducting gap $\Delta$. For $k_{\rm B} T_{\rm K} \ll \Delta$, Kondo screening is strongly suppressed on the superconducting energy scale and the impurity can be treated as an effectively static magnetic moment giving rise to well-defined YSR states. In contrast, for $k_{\rm B} T_{\rm K} \gtrsim \Delta$, quantum fluctuations and Kondo correlations become important and can qualitatively modify the impurity spectrum requiring a full quantum treatment~\cite{Huang2020Nov,Huang2023Aug,karan2022,PhysRevB.103.205424}.

In this work, we use magnetic impurities as a local probe of Ising superconductivity. We investigate the formation of YSR bound states in an Ising superconductor in the presence of an in-plane magnetic field and analyze how the interplay of ISOC, Zeeman field and impurity coupling affects the bound-state spectrum. Particular emphasis is placed on identifying features that distinguish this system from conventional superconductors and that can be accessed in tunneling experiments.

The paper is organized as follows. In Sec.~\ref{sec:model}, we introduce the theoretical model for the Ising superconductor, the magnetic impurity and their hybridization. In Sec.~\ref{sec:bound_states}, we present the resulting YSR bound-state spectrum and discuss the associated quantum phase transition. In Sec.~\ref{sec:stm}, we analyze the coupling of the YSR states to a superconducting STM tip focusing on the current-phase relation and the critical current. Finally, we summarize our results in Sec.~\ref{sec:conclusion}. Additionally, we provide a derivation of the ISC Green's function and the self-consistency equation in Apps.~\ref{sec:qcgf} and \ref{sec:selfconsistency}, a symmetry analysis for the allowed hopping matrix elements in App.~\ref{sec:symmetry} and more details on the YSR bound state computation in App.~\ref{sec:ysr_analysis} as well as a short note on the computation of the supercurrent in App.~\ref{sec:current}.

\begin{figure*}
    \centering
    \includegraphics[scale=1]{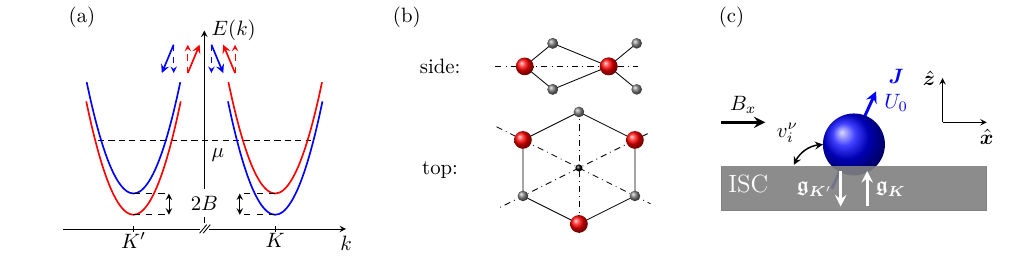}
    \caption{(a) Normal state band structure at $K,K'$ valleys with chemical potential $\mu$ and $B=\sqrt{B_x^2+\beta^2}$. The spin structure is indicated at the top of the bands for $B_x=0$ (dashed arrows) and $B_x\neq0$ (solid arrows). (b) Schematic structure of the real space lattice of 1H-TMDs. The transition metal and chalcogen atoms are depicted in red and gray respectively. The hollow site is marked by the black sphere in the center of the top view. (c) Setup of an impurity with onsite potential $U_0$ and magnetic moment $\bm{J}$ coupled to an ISC with spin-orbit coupling $\bm{\mathfrak{g}}_{\bm{k}}$ via hopping $v_i^\nu,i\in\{0,x,y,z\}$ in an in-plane magnetic field $B_x$.}
    \label{fig:setup}
\end{figure*}

\section{Model} \label{sec:model}

\subsection{Ising superconductor} \label{subsec:ISC}

We consider the normal state of a 2D TMD with spin split valence bands at the $K,K'$ points~\cite{PhysRevB.88.085433, PhysRevLett.108.196802} in the $\bm x$-$\bm y$ plane with $\bm z$ the out-of-plane direction. The normal state of these materials is well-described by the low energy effective Hamiltonian expanded around the valleys
\begin{subequations}
	\begin{align}
		H_0 &= \frac{1}{2} \sum_{\bm k} \psi^\dag_{\rm N}(\bm k)\hat h_0(\bm k) \psi_{\rm N}(\bm k) \\
		\label{eq:H0}
		\hat h_0(\bm k) &= \xi_{\bm p}\hat\sigma_0 + [\bm{\mathfrak{g}}_{\bm k} + \bm B]\cdot \hat{\bm{\sigma}}
	\end{align}
\end{subequations}
in the basis $\psi_{\rm N}(\bm k) = [c_{\bm k,\uparrow},c_{\bm k,\downarrow}]$. The total crystal momentum reads $\bm{k} = \bm{p} + \nu\bm{K}$ with the valley index $\nu=\pm$ and $\xi_{\bm p}=\bm p^2/(2m) -\mu$. Here, $\bm{\mathfrak{g}}_{\bm{k}}=\nu\beta\bm z$ denotes the valley-constant Ising spin-orbit field and $\bm B$ the external magnetic field in units of energy, absorbing a factor $g_L\mu_B/2$, where $\mu_B$ and $g_L$ respectively denote the Bohr magneton and Landé factor. The Pauli matrices $\hat\sigma_i, i\in\{x,y,z\}$ act in spin space with identity element $\hat\sigma_0$. In this work, we focus on in-plane magnetic fields as they give rise to the enhanced critical field characteristic of ISCs as mentioned in the introduction. We define the valley-dependent unitary transformation $\hat{\mathcal{U}}_\nu=\exp(i\nu\alpha\hat\sigma_y/2)$ with $\alpha=\arccos(\beta/B)$ and $B=\sqrt{B_x^2+\beta^2}$ that diagonalizes the normal state Hamiltonian according to $\hat h'_0(\bm k) = \hat{\mathcal{U}}_\nu\hat h_0(\bm k) \hat{\mathcal{U}}_\nu^\dag  = \xi_{\bm p}\hat\sigma_0 + \nu B\hat\sigma_z$. Since we assume $\bm{\mathfrak{g}}_{\bm{k}}$ to be constant at $K,K'$, this transformation does not introduce gradients and only depends on the valley index. The normal state band structure around the $K,K'$ valleys is shown in Fig.~\ref{fig:setup}.

We assume an attractive interaction producing an inter-valley s-wave pairing with mean field order parameter $\Delta\varpropto\langle c_{\bm k,\sigma} c_{-\bm k, -\sigma}\rangle$ and neglect pairing within the $\Gamma$ pocket as well as inter-valley couplings~\cite{PhysRevLett.126.237001,PhysRevB.106.184514}. The corresponding BdG Hamiltonian is block-diagonal in $\bm k$-space with blocks $\hat h_{\rm BdG}(\bm k)$ coupling electrons at $K(K')$ with holes at $K'(K)$. It reads
\begin{subequations}
	\label{eqs:HBdG}
	\begin{align}
		H_{\rm S} &= \frac{1}{2}\sum_{\bm k} \psi^\dag_{\rm S}(\bm k) \hat H_{\mathrm{BdG}} \psi_{\rm S}(\bm k), \\
		\hat H_{\mathrm{BdG}} &= \hat h_{\rm BdG}(\bm k)\oplus\hat h_{\rm BdG}(-\bm k), \\
	    \hat h_{\rm BdG}(\bm k) &= \left[ \begin{array}{cc} \hat h_0(\bm k) & \Delta i\hat\sigma_y \\ -\Delta i\hat\sigma_y & -\hat h^T_0(-\bm k) \end{array} \right],
	\end{align}
\end{subequations}
in the basis $\psi_{\rm S}^\dag(\bm k) = [c^\dag_{\bm{k},\uparrow}, c^\dag_{\bm{k},\downarrow}, c_{-\bm{k},\uparrow}, c_{-\bm{k},\downarrow}]$, where $\hat x^T$ denotes transposition of $\hat x$ in spin space. In the normal-state eigenbasis the pairing transforms as 
\[ \Delta i\hat\sigma_y\mapsto \Delta\hat{\mathcal{U}}_\nu i\hat\sigma_y\hat{\mathcal{U}}_{-\nu}^T = [\cos(\alpha)\hat\sigma_0 + i\nu\sin(\alpha)\hat\sigma_y]i\hat\sigma_y.
\]
Hence, the singlet component is reduced to $\bar\Delta=\Delta\cos(\alpha)\leq\Delta$ where the equality only holds for $B_x=0$. This explains why the observed effective gap edge in spectral plots lies below $\Delta/\Delta_0=1$ even in non-self-consistent calculations for $B_x>0$. We furthermore see that the interplay of ISOC and magnetic field produce a valley-asymmetric equal-spin triplet and momentum-independent $\bm d$-vector with $d_y\neq0$.

The normal state exhibits a spin- and valley-dependent density of states (DOS) $N_{\sigma\nu}$ with $\{\sigma\nu\}\in\{\uparrow,\downarrow\}\times\{+, -\}$ at the Fermi energy according to the Hamiltonian in Eq.~\eqref{eq:H0}. Due to spin-valley locking, only two of the four combinations are independent, i.e. $N_1 \equiv N_{\sigma\nu} = N_{\bar\sigma\bar\nu}$ and $N_2\equiv N_{\bar\sigma\nu} = N_{\sigma\bar\nu}$. In the limit of large chemical potential $\mu\gg B_x,\beta$, all bands cross the Fermi level and the difference in their DOS at the Fermi energy for different combinations $\{\sigma,\nu\}$ becomes negligible. Hence, we work with a constant, spin- and valley-independent DOS $N_0$ of the normal state in the low-energy description. The superconducting state can then be described within the quasi-classical (qc) Green's function formalism ~\cite{Kopnin2001,BELZIG19991251}. Since the BdG Hamiltonian is block-diagonal in $\bm k$-space, the quasi-classical Green's function decouples into two blocks $\hat g_\nu$. In the limit of a homogeneous and clean system with inverse Green's function $\hat g_{0,\nu}^{-1}(\varepsilon) = [\varepsilon \hat{\mathbb{1}} - \hat h_{\rm BdG}(\nu\bm K)]\hat\tau_3$, the quasi-classical Green's function follows from the Eilenberger equation $[\hat g_{0,\nu}^{-1}, \hat g_\nu] = \mathbb{0}$. In the following, we transform the Green's function $\hat g_\nu\to-i\pi N_0\hat\tau_3\hat g_\nu$ to match the commonly used notation in the community. It then admits the general form
\begin{align}
	\hat g_\nu = \pi N_0 \Big\{ &g_0 \hat\tau_0\hat\sigma_0 + g_x \hat\tau_3\hat\sigma_x + g_z \hat\tau_0\hat\sigma_z \notag\\
	\label{eq:qcgf_isc}
	&+ f_0 \hat\tau_2\hat\sigma_y + f_x \hat\tau_1\hat\sigma_z + if_y \hat\tau_1\hat\sigma_0 \Big\}
\end{align}
in the basis $\psi_{\rm S}$ with $\hat\tau_i$ denoting Pauli matrices in Nambu space with identity element $\hat\tau_0$ and six energy dependent coefficients $g_0,g_x,g_z,f_0,f_x,f_y$. Furthermore, the component functions $g_z,f_y$ are odd with respect to the valley index $\nu$. For $B_x=0,\beta\neq0$, Eq.~\eqref{eq:qcgf_isc} boils down to the Green's function of a BCS superconductor with only $g_0,f_0\neq0$. A detailed derivation of the energy and magnetic field dependent component functions can be found in App.~\ref{sec:qcgf}. The DOS in this formalism reads 
\begin{align}
	\rho = -\frac{1}{4\pi N_0} \Im(\tr(\hat g_\nu)).
\end{align}
The mean-field description of the Ising superconductor is supplemented with a self-consistent computation of the superconducting gap as a function of temperature and magnetic field as described in App.~\ref{sec:selfconsistency} whenever stated explicitly. The value at zero field and temperature is denoted by $\Delta_0$.

\subsection{YSR impurity model} \label{subsec:ysr}

In this work, we model the YSR impurity using the single-impurity Anderson model in a setup as shown in Fig.~\ref{fig:setup}. In the classical/static spin limit with explicitly lifted spin degeneracy, the model reduces to a spinful resonant level description. The Hamiltonian of the YSR impurity consists of $H_S$, the Hamiltonian of the superconducting substrate, $H_{\rm imp}$, the impurity Hamiltonian and $H_{\rm hyb}$ the hybridization between the two bare subsystems~\cite{PhysRevB.101.235445, PhysRevB.103.155407}. It reads
\begin{align}
    H_{\rm YSR} = H_S + H_{\rm imp} + H_{\rm hyb},
\end{align}
with the impurity Hamiltonian lifted to particle-hole space
\begin{subequations}
	\label{eq:imp}
	\begin{align}
		H_{\rm imp} &= \tfrac12\psi^\dag_{\rm imp} \hat h_{\rm imp}\psi_{\rm imp} \\
		\hat h_{\rm imp} &= \hat h_{{\rm imp},0}\oplus -\hat h_{{\rm imp},0}^T \\
		\label{eq:himp}
		\hat h_{{\rm imp},0} &= U_0\hat\sigma_0 + \bm B_{\rm imp}\cdot\hat{\bm{\sigma}}
	\end{align}
\end{subequations}
in the basis $\psi_{\rm imp}=[d_\uparrow,d_\downarrow,d^\dag_\uparrow,d^\dag_\downarrow]$. We include an on-site potential $U_0$ and a fixed magnetic moment $\bm J$, respectively breaking particle-hole symmetry (PHS) and spin-symmetries, locally on the impurity. In our model, the magnetic field enters as a static Zeeman term acting locally on the impurity and in the superconductor. In principle it also enters together with the local exchange interaction in the gap equation at the impurity site leading to a qualitative change in the behaviour as discussed in Ref.~\cite{PhysRevB.111.224501}. In this work, we neglect this effect and instead introduce the total local magnetic moment on the impurity $\bm B_{\rm imp}\equiv\bm B + \bm J$. Different to YSR states in BCS superconductors, we need to explicitly consider the direction of the impurity moment as the ISC is not spin-rotation invariant. In the following, we restrict attention to the subspace of impurity realizations satisfying $\bm J \cdot \bm y = 0$, i.e. lying in the plane spanned by the in-plane magnetic field and the ISOC. We parametrize the impurity’s local moment by $\bm J = J[\sin(\theta), 0, \cos(\theta)]$. The projection onto the impurity spin then reads $\hat R=\exp(i\vartheta\hat\sigma_y/2)$ with $\vartheta=\arctan([J \sin(\theta) + B_x] / [J\cos(\theta)])$. 
For magnetic moments $J$ similar to the external field or the ISOC strength, we expect a non-negligible electronic feedback on the impurity moment and thus a qualitative change in the behaviour of the impurity also depending on the actual magnitude of Coulomb interaction on the impurity. However, for large Coulomb repulsion compared to the hybridization strength, $U\gg\Gamma$ this effect is expected to be small and the fixed moment description remains valid. The bare impurity Green's function follows by $\hat g_{\rm imp}(\varepsilon) = [\varepsilon \hat{\mathbb{1}} - \hat h_{\rm imp}]^{-1}$. Throughout this work, we assume the magnetic adatom to be positioned at the hollow site in the real space lattice as indicated in Fig.~\ref{fig:setup}.

\subsubsection{Hybridization with substrate} \label{sec:hybridization}

The impurity is coupled to the substrate by the particle number conserving hybridization Hamiltonian
\begin{align}
    \label{eq:coupling}
	H_{\rm hyb} = \sum_{\bm k} \psi^\dag_{\rm S}(\bm k) \hat V(\bm k)\psi_{\rm imp} + \text{h.c.}.
\end{align}
In the presence of a uniaxial in-plane field $B_x$ and a generic impurity moment $\bm J$ with $J_z\neq 0$, time-reversal symmetry, threefold rotation symmetry $C_3$ and in-plane mirror symmetry $\sigma_h$ are broken. Consequently, the spin-structure of the hopping matrix is only restricted by hermiticity. To simplify the computation, we assume a smooth impurity potential $U(\bm r)$ with characteristic width $\xi$. Its Fourier transform $U(\bm q)$ is strongly suppressed for momentum transfers $|\bm q|\gtrsim \xi^{-1}$. Since inter-valley processes between states near $\bm K$ and $\bm K'$ require momentum transfer $\bm q \approx\bm K - \bm K'$, they are negligible provided $1/\xi\ll |\bm K-\bm K'|$. For a hexagonal lattice this amounts to $\xi \gg 3a/(4\pi)$. We therefore project onto the low-energy valley subspace
\begin{align*}
	\mathcal H_{\rm eff} = \bigoplus_{\nu=\pm}\left\{ \ket{\nu,\bm k}\colon |\bm k-\nu\bm K|\ll \xi^{-1}\right\},
\end{align*}
and retain only valley-diagonal impurity matrix elements to leading order. Note that valley-diagonal does not imply valley-symmetric as the projected couplings may still depend on $\nu$. As outlined in App.~\ref{sec:symmetry}, the normal state hopping matrix admits the general form
\begin{align}
	\label{eq:hoppingmatrix}
	\hat V(\nu\bm K) = v_0^\nu \hat\sigma_0 + \bm v^\nu\cdot\hat{\bm{\sigma}}
\end{align}
with eight independent functions $v_i^\nu$ which up to lowest order are assumed to be constant in $\bm k$ around the valleys. To simplify the analysis, we focus on three different scenarios of hopping matrices with non-zero elements $v_1,v_2$ according to:
\begin{enumerate}
	\item[(i)] spin-polarized hopping, valley-symmetric,
		\[ v^\pm_0=v_1, v^\pm_z=v_2\]
	\item[(ii)] spin-flip hopping, valley-symmetric, 
		\[v^\pm_x=v_1, v^\pm_y=v_2\]
	\item[(iii)] spin-independent hopping, valley-asymmetric,
		\[v_0^+ = v_1, v_0^-=v_2.\]
\end{enumerate}

The standard BCS result is obtained using $v_2=0$ in case (i) also leading to a single tunneling rate $\Gamma=\pi N_0 |v_1|^2$ which motivates our choice of plotting parameter in the bound state spectrum. The YSR impurity is then described by the dressed Green's function following from the Dyson equation
\begin{align}
    \label{eq:isingysr}
    \hat G_{\rm imp} &= \hat g_{\rm imp} + \hat g_{\rm imp}\hat\Sigma_S \hat G_{\rm imp},
\end{align}
with $\hat\Sigma_S = \tfrac12\sum_\nu\hat W_\nu\hat g_\nu\hat W_\nu^\dag$ and the hopping matrix $\hat W_{\nu} = \text{diag}(\hat V(\nu\bm K), -\hat V^*(\bar\nu\bm K))$ in Nambu space. Here, the factor $\tfrac12$ accounts for the valley doubling. We want to highlight, that since $g_z,f_y$ in the ISC are odd in valley, they only contribute to the self-energy in case (iii).

\subsubsection{YSR bound states} \label{sec:ysr_bs}

\begin{figure}[ht]
	\centering
	\includegraphics[scale=1]{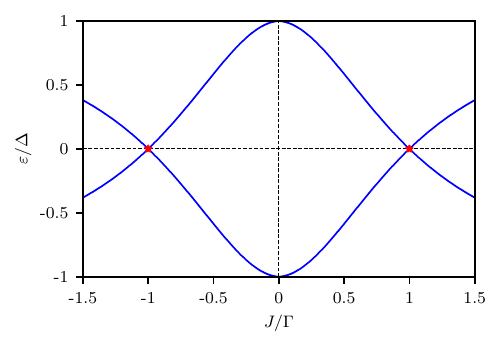}
	\caption{Bound state energy of BCS YSR states for $\{\Gamma,U_0\}=\{50,0\}\Delta_0$ and values of $J_{\rm eff}=0$ indicated by red stars.}
	\label{fig:bcsysr}
\end{figure}

We extract the bound states from the inverse Green's function
\begin{align}
	\hat G_{\rm imp}^{-1} = \hat g_{\rm imp}^{-1} - \hat\Sigma_S,
\end{align}
as the solutions of $\det\,(\hat G_{\rm imp}^{-1}) = 0$. For the three cases of hybridization, the substrate self-energy $\hat\Sigma_S(\varepsilon)$ acquires different energy-dependent structures in spin-Nambu space. To isolate the relevant part of the spectrum, we project onto the electron ($\tau=1$) and hole ($\tau=-1$) blocks. Provided that the opposing block is invertible, i.e. $\det\big([\hat G_{\rm imp}^{-1}]^{\bar\tau\bar\tau}\big)\neq0$, a Schur-complement reduction yields
\begin{align}
	[\hat G_{\rm imp}^{\tau\tau}]^{-1} = \mu_\tau\hat\sigma_0 + \bm J^\tau_{\rm eff}\cdot\hat{\bm{\sigma}} + i\hat\Gamma
\end{align}
for each $\tau$. Here, we have introduced an effective scalar potential $\mu_\tau\in\mathbb R$ and an effective magnetic moment $\bm J^\tau_{\rm eff}(\varepsilon)\equiv\Re\big(\bm B_{\rm imp}+\bm B_\tau(\varepsilon)\big)\in\mathbb R^3$ as well as a spin-dependent broadening $\hat\Gamma\in\mathbb R^{2\times2}$. The energy-dependent spin splitting $\bm B_\tau(\varepsilon)$ originates from off-diagonal terms in spin-space of the self-energy and contains contributions from pairing correlations in the ISC through the Schur complement as outlined in App.~\ref{sec:bound_states_app}. More explicitly, for $B_x\neq0$ it includes terms proportional to the triplet components $f_x$ and $f_y$ not present in the BCS case. In the subgap regime and for $\eta\to0$, bound states at $\varepsilon=\varepsilon_0$ correspond to poles on the real energy axis and satisfy the condition
\begin{align}
	\mu_\tau(\varepsilon_0) = \pm|\bm J^\tau_{\rm eff}(\varepsilon_0)|.
\end{align}

Since by construction the BdG Hamiltonian in Eq.~\eqref{eqs:HBdG} is particle-hole symmetric (PHS) and our choice of hopping matrices preserves this symmetry, also the YSR Green's function fulfils a PHS relation. In the regime, where the Schur complement reduction is valid, the effective magnetic moments in the two $\tau$-sectors are related by
\begin{align}
	\label{eq:j_phs}
	\bm J^{\rm e}_{\rm eff}(\varepsilon_0) = -\bm J^{\rm h}_{\rm eff}(-\varepsilon_0) ,
\end{align}
implying a mirror symmetry of the bound-state energies about $\varepsilon=0$, even for explicitly broken PHS on the impurity by $U_0\neq0$. For a spin-rotation invariant substrate the $\tau$-block description is redundant and one always finds only a single particle-hole related pair of quasiparticle states. More explicitly, this is the case for YSR states on standard BCS superconductors as we show in Fig.~\ref{fig:bcsysr}. More details on the calculation for a BCS substrate can be found in App.~\ref{par:bcs_example}. If the substrate breaks spin-rotation symmetry, one instead finds two pairs of bound states. One pair crosses at $\varepsilon=0$ and marks the quantum phase transition of the YSR problem~\cite{karan2022}. Additionally, depending on the parameters of the problem, another crossing or anti-crossing of the bound states occurs at finite energy. The crossings are protected by an internal symmetry whenever the corresponding effective Hamiltonians $\hat h_{\rm eff}^\tau = \varepsilon_0\mathbb 1 - [\hat g_{\rm YSR}^{\tau\tau}]^{-1}(\varepsilon_0)$ are simultaneously diagonalizable for both $\tau$ and satisfy $|\bm J^{\rm e}_{\rm eff}\times\bm J^{\rm h}_{\rm eff}|=0$. Further details of this symmetry analysis are provided in App.~\ref{sec:bs_symmetry}.

\subsubsection{Supercurrent}

We study the Josephson current between a superconducting STM tip and the YSR impurity by solving the Dyson equation for the dressed impurity Green's function $\hat G_{\rm imp}(\varphi)$ in the presence of tip-induced self-energy corrections $\hat\Sigma_t = V_t \hat g_t V_t^\dag$ according to
\begin{align}
	\hat G_{\rm imp}(\varphi) = \hat g_{\rm imp} + \hat g_{\rm imp} \hat\Sigma \hat G_{\rm imp}(\varphi).
\end{align}
with $\hat\Sigma=\hat\Sigma_t + \hat\Sigma_S$. The superconducting tip is modelled as a homogeneous spin-singlet BCS superconductor with constant normal-state density of states $N_{0,{\rm BCS}}$ at the Fermi energy and phase difference $\varphi$ relative to the substrate. For reasons of simplicity, we choose a momentum- and spin-independent hopping amplitude $t_t$ leading to hopping matrix $V_t = t_t \hat\tau_3\hat\sigma_0$ as well as the tunneling rate $\Gamma_t\equiv\pi N_{0,{\rm BCS}}|t_t|^2$. In the basis $\psi_t = [a_{\bm k\uparrow},a_{\bm k\downarrow},a^\dag_{-\bm k\uparrow},a^\dag_{-\bm k\downarrow}]$, the quasi-classical Green's function of the tip reads
\begin{equation}
    \label{eq:tip}
    \hat g_t(\varepsilon) = -\pi N_{0,{\rm BCS}} \frac{\varepsilon \hat\tau_0\hat\sigma_0 - \Delta e^{i\varphi\hat\tau_3\hat\sigma_0}\hat\tau_2\hat\sigma_y}{\sqrt{|\Delta|^2 - \varepsilon^2}}.
\end{equation}

The equilibrium supercurrent follows from the phase dependence of the dressed impurity Green's function and can be written in terms of the tip self-energy as
\begin{equation}
    \label{eq:supercurrent}
    I(\varphi) = \frac{e}{\hbar}\int_{\mathbb R} d\varepsilon \Re\left[\tr\, \left(\hat\tau_3[\hat\Sigma_t^a(\varepsilon),\hat G_{\rm imp}^a(\varepsilon)]\right)\right] f(\varepsilon),
\end{equation}
where $f(\varepsilon)$ is the Fermi distribution and the superscript $a$ denotes the analytic continuation $\varepsilon\to\varepsilon - i0^+$. For the singlet tip considered here, the supercurrent probes the anomalous components $G_{1y}\equiv\tr\,(\hat\tau_1\hat\sigma_y \hat G_{\rm imp})/4$ and $G_{2y}\equiv\tr\,(\hat\tau_2\hat\sigma_y \hat G_{\rm imp})/4$ which contain the information about the impurity-induced subgap spectrum. For reasons of simplicity we set $N_{0,{\rm BCS}}=N_0$ in the following.

\section{YSR Bound state spectrum}
\label{sec:bound_states}
\begin{figure*}[ht]
	\centering
	\includegraphics[scale=1]{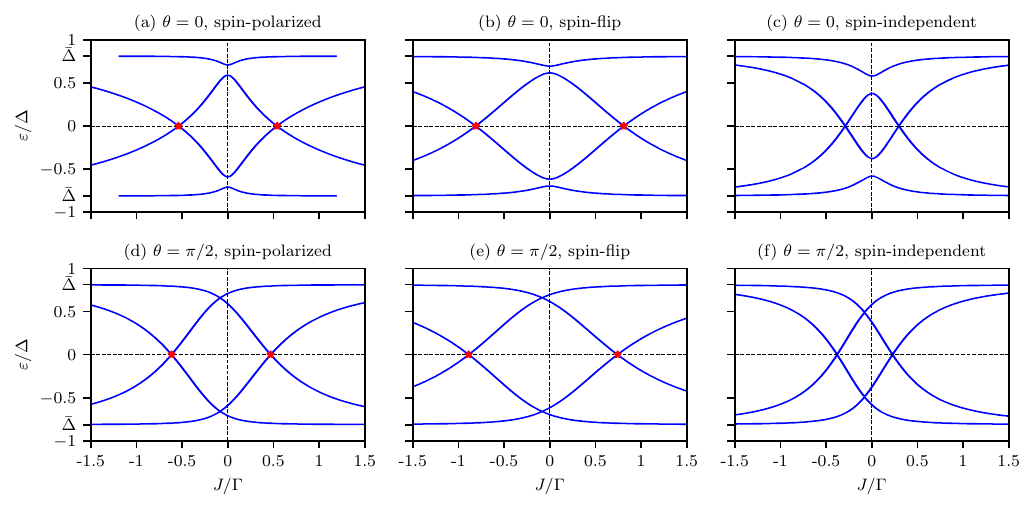}
	\caption{Bound-state energies for $\{\beta,B_x,U_0,\pi N_0|v_1|^2\}=\{7,5,0,50\}\Delta_0$, $|v_1|^2/|v_2|^2=5$, $\theta=0,\pi/2$ in (a)--(c), (d)--(f), respectively, and the indicated hopping cases. The gap is calculated self-consistently for $T=10^{-3}T_c$. Red stars indicate symmetry-protected state crossings.}
	\label{fig:ysrstates}
\end{figure*}
The aim of this section is to highlight the differences in the bound-state spectrum of YSR states on an ISC substrate compared to the corresponding spectrum on a conventional BCS substrate, which is shown in Fig.~\ref{fig:bcsysr} as a reference. For the ISC substrate, we present the YSR bound-state spectrum as a function of the bare impurity magnetic moment $J$ for the three hopping matrices introduced before in Fig.~\ref{fig:ysrstates}. The impurity orientation is characterized by the angle $\theta$ with respect to the ISOC direction in the $x$--$z$ plane, with $\theta=0$ and $\theta=\pi/2$ shown in the top and bottom rows, respectively. 

The bound-state energies are computed in the strong-coupling regime with $\Gamma_1 = \pi N_0 |v_1|^2 = 50\Delta$ and $\Gamma_2 = \pi N_0 |v_2|^2 = 10\Delta$, as well as strong Ising spin-orbit coupling $\beta = 7\Delta$ and an in-plane magnetic field $B_x = 5\Delta$. The persistence of YSR states up to such large in-plane magnetic fields is solely due to Ising protection in the substrate and constitutes a hallmark feature by itself~\cite{PhysRevLett.126.237001,PhysRevB.106.184514}.

Compared to YSR states in a BCS substrate, two main differences arise. First, in the ISC case, we observe two distinct subgap states for $|\varepsilon|\leq\bar{\Delta}$ that are symmetric around $\varepsilon=0$. Depending on the choice of $\theta$ and the hopping matrix, a gap at finite energy may open in the bound-state spectrum. Second, there is an explicit dependence on the direction of the bare impurity magnetic moment. This effect is most pronounced around $J/\Gamma=0$ in Fig.~\ref{fig:ysrstates} (d)--(f). For $\theta=0$, the bound-state spectrum is symmetric under $J \to -J$, whereas for $\theta=\pi/2$, it becomes asymmetric and the finite-energy crossing is shifted to the left for $B_x > 0$. 

These features directly result from the simultaneous action of the in-plane magnetic field on both the impurity and the substrate. For $\theta=\pi/2$, collinear and anti-collinear configurations of the magnetic moment and the in-plane field lead to different magnitudes $|\bm{B}_{\rm imp}|$, which also explains the offset at $J/\Gamma=0$.

The explicit dependence on the direction of the impurity magnetic moment and the doubling of the number of YSR states discussed above originate from the combined effect of an in-plane magnetic field and an out-of-plane ISOC, which explicitly breaks spin-rotation symmetry within each $\nu$-block of Eq.~\eqref{eqs:HBdG}. To see this, it is useful to recall the spin structure of the normal state in 2D TMDs subject to an in-plane magnetic field, schematically indicated by the dashed arrows in Fig.~\ref{fig:setup}(a). Since superconductivity is introduced such that $\ket{\uparrow,K}$ and $\ket{\downarrow,K'}$ are paired at the Fermi surface, the magnetic field generates a finite spin component along the field direction. This selects a preferred spin axis for each combination in the BdG Hamiltonian of Eq.~\eqref{eqs:HBdG} and thus breaks spin-rotation invariance.

For valley-asymmetric hopping, we furthermore observe a gap widening of the gap in the spectrum for $\theta=0$ as well as the absence of symmetry protected crossings in Fig.~\ref{fig:ysrstates} (c), (f). This is caused by the valley-antisymmetric components $g_z$ and $f_y$ in $\bm B_\tau$. Under the combined transformation $\varepsilon\to -\varepsilon$ and $\tau\to\bar{\tau}$, these terms do not preserve the simple particle-hole symmetry relation in Eq.~\eqref{eq:j_phs} at the bound-state energy, as discussed in App.~\ref{sec:effective_field}.
\begin{figure}[!b]
	\centering
	\includegraphics[scale=1]{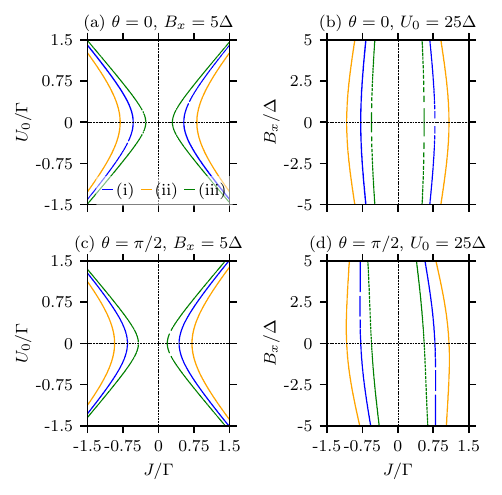}
	\caption{QPT condition from Eq.~\eqref{eq:qpt_cond} as a function of onsite potential $U_0$ in (a),(c) and magnetic field strength $B_x$ in (b),(d) for the indicated hopping cases, with $\{\beta,\pi N_0|v_1|^2\}=\{7,50\}\Delta$, $|v_1|^2/|v_2|^2=5$, and impurity realizations $\theta=0,\pi/2$ in (a),(b) and (c),(d).}
	\label{fig:qpt_cond_ISC}
\end{figure}

\paragraph*{Quantum phase transition.} 
In general, a quantum phase transition (QPT) in YSR systems occurs when the lowest-energy states in the even- and odd-parity sectors of the impurity many-body spectrum cross. In the quasiparticle spectrum shown in Fig.~\ref{fig:ysrstates}, this transition therefore appears naturally at $\varepsilon=0$. To describe these states in ISCs, we derive an effective Hamiltonian from the on-shell inverse Green's function, rotated onto the impurity spin-quantization axis and evaluated at the bound-state energy. In contrast to YSR states on a spin-rotation-invariant substrate, this rotation must be performed after computing the dressed YSR Green's function, since the substrate couples electronically to the impurity moment and thereby modifies the spin structure.

After integrating out the substrate degrees of freedom, the impurity is described by the quadratic Hamiltonian
\begin{align}
	H_{\rm eff}\equiv\sum_\sigma \varepsilon_\sigma d^\dag_{\sigma}d^{\phantom{\dag}}_\sigma + \Gamma_{\uparrow\downarrow} d_\uparrow d_{\downarrow} + \Gamma^*_{\uparrow\downarrow} d^\dag_{\downarrow}d^\dag_\uparrow,
\end{align}
which breaks particle-number conservation through the induced pairing term $\Gamma_{\uparrow\downarrow}$ while preserving fermion parity. The many-body Hilbert space of a single level, $\mathcal H_{\rm mb}=\{\ket{0},\ket{\uparrow},\ket{\downarrow},\ket{\uparrow\downarrow}\}$, then separates into even- and odd-parity sectors. Projecting onto these sectors yields the corresponding many-body states and energies. As shown in App.~\ref{sec:qpt}, the crossing of the two lowest states leads to the general QPT criterion
\begin{align}
	\label{eq:qpt_cond}
	|\Gamma_{\uparrow\downarrow}|^2 + \varepsilon_\uparrow\varepsilon_\downarrow = 0.
\end{align}
In Fig.~\ref{fig:qpt_cond_ISC}, we investigate this condition as a function of the onsite potential $U_0$ at fixed $B_x$ in (a), (c) and as a function of $B_x$ at fixed $U_0$ in (b), (d). The two impurity orientations $\theta=0$ and $\theta=\pi/2$ are considered in (a), (b) and (c), (d), respectively. All remaining parameters are identical to those used in the bound-state spectrum calculation.

\begin{figure*}[!t]
    \centering
	\includegraphics[scale=1]{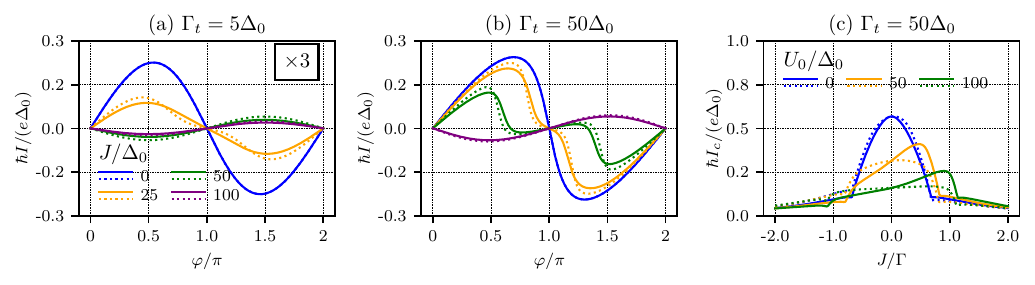}
	\caption{(a),(b) CPR for different $\Gamma_t$ and $J$ as indicated in the legend. (c) Critical current for $U_0$ as indicated in the legend with $\Gamma=\pi N_0(|v_1|^2+|v_2|^2)+\Gamma_t$. We use $\{\beta, B_x, \pi N_0|v_1|^2, U_0, \eta\} = \{7, 5, 50, 0, 0.01\}\Delta_0$, $|v_1|^2/|v_2|^2 = 5$ and $T=0.01 T_{c,0}$. Solid, dotted lines denote correspond to $\theta = 0$, $\theta = \pi/2$, respectively.}
    \label{fig:current}
\end{figure*}

Compared with the BCS case, where the QPT condition reads $J/\Gamma = \sqrt{1 + [U_0/\Gamma]^2}$, the minimal value of $J/\Gamma$ at which a QPT can occur is substantially reduced, depending on the hopping matrix under consideration. We furthermore find the same asymmetry under $J\to -J$ for $\theta=\pi/2$ as in the bound-state spectrum, and for the same reason, also under $B_x\to -B_x$. Reversing the signs of $J$ and $B_x$ simultaneously leaves the QPT point unchanged, which appears as point symmetry around the origin in Fig.~\ref{fig:qpt_cond_ISC} (d).

\section{Supercurrent to a BCS Tip and Andreev Bound States} \label{sec:stm}

\begin{figure}[ht]
	\centering
	\includegraphics[scale=1]{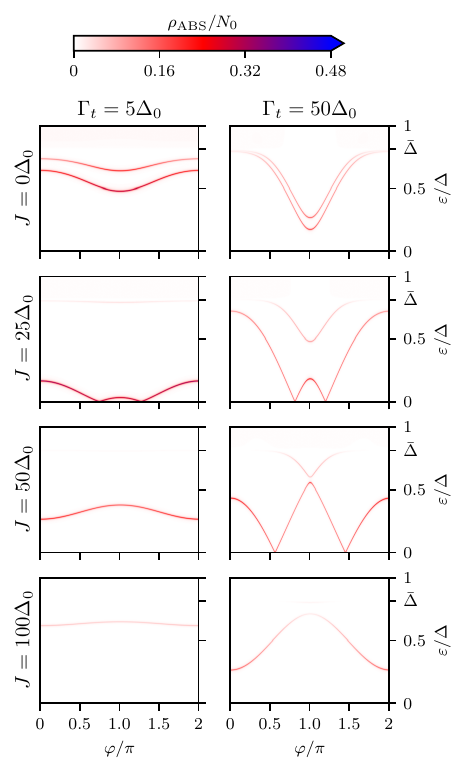}
	\caption{LDOS of ABS over the phase difference $\varphi$ for $\{\beta, B_x, \pi N_0 |v_1|^2, U_0\} = \{5, 7, 50, 0\}\Delta_0$, $|v_1|^2/|v_2|^2 = 5$, $\theta = \pi/2$ and indicated values of $J$ for hopping case (i). The superconducting gap is computed self-consistently for $T=0.01T_{c,0}$ and Dynes broadening $\eta=0.01\Delta_0$.}
	\label{fig:ysrabs}
\end{figure}

In this section, we analyze the equilibrium supercurrent flowing through the YSR impurity in a junction formed with a BCS scanning tunneling microscope (STM) tip. We show in Fig.~\ref{fig:current}(a,b) the current-phase relation (CPR) for different tunneling rates $\Gamma_t$ to the tip, impurity realizations $\theta$ and values of the impurity magnetic moment $J$. For the same set of parameters, Fig.~\ref{fig:ysrabs} displays the corresponding positive energy branch of the Andreev bound-state (ABS) dispersion in the weak-coupling (left column) and the strong-coupling (right column) limit for the impurity orientation $\theta=\pi/2$. This direct comparison allows us to relate the current obtained from Eq.~\eqref{eq:supercurrent} arising from the phase-dependent hybridization between the tip and the YSR state to the more intuitive picture
\begin{align}
	I(\varphi) = -\frac{2e}{\hbar} \sum_{E_n>0} f(E_n) \partial_\varphi E_n(\varphi),
\end{align}
based on the phase dispersion of the ABS energies $E_n$. On top of the dispersion of the ABS, the CPR also reflects the occupation of the ABS via the Fermi function $f(E)$. As in the case without the STM tip, we observe two distinct subgap states contributing to the supercurrent to the tip. We distinguish between weak- and strong-coupling regimes in terms of the ratio of tunneling rates $\Gamma_t/\Gamma_S$. 

In the weak-coupling limit $\Gamma_t\ll\Gamma_S$, corresponding to the typical STM regime, the Josephson current is governed primarily by the impurity-induced subgap spectrum and the CPR remains close to sinusoidal away from the QPT. In this regime, the ABS exhibit only a weak phase dispersion which becomes progressively smaller as the magnetic moment $J$ increases as displayed in the left column of Fig.~\ref{fig:ysrabs} from top to bottom. For $J=0$, corresponding to the non-magnetic case shown in the top left panel of Fig.~\ref{fig:ysrabs}, the ABS display the largest phase dispersion and thus yield the largest critical current $I_c=\max_{\varphi} I(\varphi)$. Upon increasing $J$, the phase dispersion is reduced and the amplitude of the CPR decreases accordingly in Fig.~\ref{fig:current}(a). In the vicinity of the QPT, the CPR undergoes a qualitative change as higher harmonics become significant and $\partial_\varphi E_n$ changes sign multiple times. Since the QPT occurs at different values of $J$ for the two impurity realizations, the resulting CPRs differ significantly near the transition.

The pronounced non-sinusoidal response follows from a rapid change in the occupation of the bound states as they approach the Fermi level. This is accompanied by a strong redistribution of spectral weight within the subgap sector, which generates higher-harmonic contributions. Across the QPT, this results in the characteristic $0$ to $\pi$ transition, where the supercurrent reverses sign, as already known from conventional BCS YSR systems~\cite{karan2022,Kuster2021}. Beyond the transition, the current is carried mainly by the deep subgap state since the other state is pushed towards the continuum and disperses less with $\varphi$. As the remaining state is pushed closer to the gap edge, the critical current decreases further and the CPR becomes strongly suppressed. Although the two impurity realizations remain distinguishable in principle for all values of $J$, their differences become negligible once the overall current is small. 

In the strong-coupling regime, the ABS in the right column of Fig.~\ref{fig:ysrabs} develop a much stronger phase dispersion resulting in a larger critical current as well as a non-sinusoidal behaviour for values of $J$ below the QPT or equivalently below the $0$ to $\pi$ transition as displayed in Fig.~\ref{fig:current}(b). This behaviour reflects a general feature of tunnel-coupled Josephson junctions where the largest deviations from the $\sin(\varphi)$ behaviour occur in the vicinity of the quantum phase transition of the dressed impurity state. In our case this is accompanied by the largest difference between impurity realizations near the QPT.

In Fig.~\ref{fig:current}(c), we show the critical current for the same set of parameters, but for different values of the onsite potential $U_0$ as indicated in the legend. We observe substantial variations of the critical current between different impurity realizations, as well as the same asymmetry under $J \to -J$ for $\theta=\pi/2$. This asymmetry can also be observed for fixed $J$ and a reversal of the in-plane magnetic field $B_x\to-B_x$. Under certain conditions, e.g. a large on-site potential, the critical current is strongly suppressed for the anti-collinear field configuration at the qpt even for $T\neq0$ in the strong coupling regime. This trend can already be seen for small temperatures $T=0.001T_{c,0}$ in the large $U_0$ plot of Fig.~\ref{fig:current}(c).

\section{Conclusion} \label{sec:conclusion}

In this work, we have investigated magnetic impurities coupled to Ising superconductors within the single-impurity Anderson model in the classical-spin limit under the influence of an in-plane magnetic field. Motivated by the ongoing debate on the nature of the pairing state in two-dimensional transition-metal dichalcogenides, our goal was to identify robust spectral signatures that directly reflect the spin structure of the underlying superconducting condensate. To this end, we analyzed a range of symmetry-allowed hybridization channels and demonstrated that the resulting Yu-Shiba-Rusinov (YSR) spectrum acquires a rich and characteristic structure due to the interplay of Ising spin-orbit coupling and the in-plane Zeeman field. The absence of spin-rotation symmetry in the substrate leads in general to a splitting of the impurity-induced subgap spectrum into two spin-dependent branches. We have shown that the corresponding YSR energies depend sensitively on the relative orientation between the impurity moment and the applied field. For non-collinear configurations, the two branches hybridize giving rise to avoided crossings and the opening of finite-energy gaps in the spectrum. Furthermore, we find that the quantum phase transition deviates significantly from the conventional BCS YSR estimate and can occur at substantially reduced ratios of magnetic moment and tunneling rate. Its precise location depends on the hybridization channel as well as on the magnetic field, the impurity onsite potential and the strength of the Ising spin-orbit coupling. These spectral features translate directly into measurable signatures in tunneling experiments. In particular, for a superconducting STM probe, we predict a pronounced dependence of the critical current on both the impurity onsite potential and the orientation of the impurity moment. This sensitivity establishes magnetic impurities as versatile local probes of the spin structure of Ising superconductors and provides experimentally accessible fingerprints that distinguish them from conventional superconducting states.

Our results therefore support the use of impurity-based spectroscopy as an additional tool to resolve the pairing structure in candidate Ising superconductors. An interesting direction for future work is the use of functionalized or spin-polarized STM tips, which allow to access spin-resolved transport signatures and further sharpen the connection between impurity-induced bound states and the underlying superconducting order.

\section{Acknowledgements}

M.H. would like to thank Gaomin Tang for valuable input on the numerical computation. M.H. and W.B. acknowledge support by the Deutsche Forschungsgemeinschaft (DFG; German Research Foundation) via Project No. 465140728, Project No. 443404566 and SFB 1432 (Project No. 425217212). J.C.C. thanks the Spanish Ministry of Science and Innovation (contract no.\ PID2024-157536NB-C22) and the ``Maria de Maeztu" Programme for Units of Excellence in R\&D (CEX2023-001316-M).

\appendix

\section{Quasi-classical Green's function of ISC}
\label{sec:qcgf}
We compute the quasi-classical Green's function $\hat g'_\nu$ of the ISC in the basis $\psi'_{\rm S}$ from the main text, where the normal state is diagonal. The quasiparticle bands follow from the energy eigenvalues of $\hat h_{\rm BdG}(\nu\bm k)$ as
\begin{align}
	E(\bm k) = \pm\sqrt{\xi_{\bm p}^2 + B^2 + \Delta^2 \pm 2 B \sqrt{\xi_{\bm p}^2 + \Delta^2s_\alpha^2}}.
\end{align}
Here and below, we use the shorthand notation
\begin{align}
	s_\alpha\equiv\sin(\alpha)\quad\text{and}\quad c_\alpha\equiv\cos(\alpha).
\end{align}
We introduce the 16 basis matrices $\hat E_{ij}\equiv \hat\tau_i\otimes\hat\sigma_j$ of $\text{Mat}(4,\mathbb{C})$ with $i\in\{0,1,2,3\}$ and $j\in\{0,x,y,z\}$. Apart from terms $\sim E_{00}$, the inverse Green's function reads
\begin{align}
	\label{eq:A}
	\hat A\equiv\varepsilon \hat E_{30} - \nu B\hat E_{3z} + i\Delta c_\alpha \hat E_{1y} - i\nu\Delta s_\alpha \hat E_{20}.
\end{align}
The BdG nature of $\hat h_{\rm BdG}(\nu\bm k)$ in Eq.~\eqref{eqs:HBdG} of the main text leads to a particle-hole symmetric (PHS) spectrum $\text{spec}(\hat A)=\{\lambda_+,\lambda_-,-\lambda_+,-\lambda_-\}$ with
\begin{align}
	\lambda_\pm=\sqrt{\varepsilon^2 + B^2 - \Delta^2 \pm 2B\sqrt{\varepsilon^2-\Delta^2 c_\alpha^2}}.
\end{align}
The characteristic polynomial reads
\begin{align}
	\chi(z) = z^4 - (\lambda_+^2+\lambda_-^2)z^2 + \lambda_+^2\lambda_-^2 \in \mathbb C[z^2],
\end{align}
where $\mathbb C[z^2]$ denotes the algebra of even polynomials. The qc gf is determined by $[\hat A,\hat g'_\nu]=\mathbb 0, \hat g'^2_\nu=\hat{\mathbb{1}},\tr \hat g'_\nu=0$. We select the retarded solution by analytic continuation $\varepsilon\to\varepsilon+i0^+$, which fixes the branch of the square root in the upper half-plane. For generic parameter values away from isolated degeneracy points, $\hat A$ has a non-degenerate spectrum and is diagonalizable. Hence, any matrix commuting with $\hat A$ is a polynomial in $\hat A$. In particular for $\lambda_\pm\neq0,\lambda_+^2\neq\lambda_-^2$, we have $\hat g_\nu'\in\mathbb C[\hat A]$ and by Cayley-Hamilton, we may write
\begin{align*}
	\hat g_\nu'=p(\hat A), \qquad p\in\mathbb C[z], \deg(p)\leq 3.
\end{align*} 
We aim to find $p(\hat A) = \sum_{j=1}^4 s_j L_j(\hat A)$ using the Lagrange polynomials $L_j(\lambda_k)=\delta_{jk}$
\begin{align}
	L_j(z) = \prod_{m\neq j} \frac{z-\lambda_m}{\lambda_j-\lambda_m}
\end{align}
and $s_j\equiv\lim_{\eta\to0^+}\text{sign}(\Im(\lambda_j(\varepsilon+i\eta)))$. Since $s(-\lambda_i)=-s(\lambda_i)$, $p(z)$ is necessarily an odd polynomial. The coefficients in $p(z)=\sum_{k=0}^3 c_k z^k$ follow by
\begin{align}
	c_k = \sum_{j=1}^4 s_j S_k^{(j)} \prod_{m\neq j}\frac{1}{\lambda_j-\lambda_m}
\end{align}
with the odd power coefficients
\begin{align}
	S_1^{(j)}\equiv \sum_{\substack{m<n\\ m,n\neq j}}\lambda_m\lambda_n, \qquad S_3^{(j)}\equiv1.
\end{align}
More explicitly, we find $p(z)=z(u+vz^2)$ with
\begin{align}
	v = \frac{s_+\lambda_- - s_-\lambda_+}{\lambda_+\lambda_-[\lambda_+^2-\lambda_-^2]} \quad\text{and}\quad u = \frac{s_+}{\lambda_+} - v\lambda_+^2
\end{align}
and thus $\hat g'_\nu = u \hat A + v\hat A^3$. According to Cayley-Hamilton, the algebra $\mathbb C[\hat A]$ carries a natural $\mathbb Z_2$-grading, i.e. $\mathbb C[\hat A] = \mathbb C[\hat A]_{\rm even} + \mathbb C[\hat A]_{\rm odd}$ with
\begin{alignat*}{3}
	&\mathbb C[\hat A]_{\rm even} &&= \spanof{\hat{\mathbb{1}},\hat A^2}, \\ 
	&\mathbb C[\hat A]_{\rm odd} &&= \spanof{\hat A,\hat A^3}
\end{alignat*}
as expected for a Hamiltonian with PHS. We have
\begin{align}
	\hat A^3 = &-B\Delta^2s_{2\alpha} \hat E_{0x} + \varepsilon[3B^2+\varepsilon^2-\Delta^2]\hat E_{30} \notag\\
	&+ \nu B[\Delta^2(1+2c_\alpha^2)-B^2-3\varepsilon^2] \hat E_{3z} \notag\\
	& +i\Delta c_\alpha[3B^2+\varepsilon^2-\Delta^2]\hat E_{1y} \notag\\
	& -i\nu \Delta s_\alpha [B^2+\varepsilon^2-\Delta^2] \hat E_{20} \notag \\
	\label{eq:A3}
	&+ 2iB\varepsilon\Delta s_\alpha \hat E_{2z}
\end{align}
and $\hat g'_\nu\in\spanof{\hat E_{0x},\hat E_{30},\hat E_{3z},\hat E_{1y},\hat E_{20},\hat E_{2z}}$. Note that the $v \hat A^3$ contribution generates additional matrix structures corresponding to triplet correlations. So far, we have computed $\hat g'_\nu$ in the basis $\psi_{\rm S}'$. In the original basis we have $\hat g_\nu=\hat R_\nu^\dag\hat g'_\nu\hat R_\nu$ with
\begin{align}
	\hat R_\nu = \hat{\mathcal{U}}_\nu\oplus\hat{\mathcal{U}}^*_{-\nu}=c_{\alpha/2}\hat E_{00} + i\nu s_{\alpha/2}\hat E_{3y}.
\end{align}
The basis matrices transform according to
\begin{align}
	\label{eq:mat_transform}
	\hat E_{ij}\mapsto\hat R_\nu^\dag \hat E_{ij}\hat R_\nu = &[c_{\alpha/2}^2 + s_{\alpha/2}^2 \kappa_{ij}] \hat E_{ij} \notag \\
	&+i\nu c_{\alpha/2}s_{\alpha/2} [1-\kappa_{ij}]\hat E_{ij}\hat E_{3y}
\end{align}
with $\hat E_{3y}\hat E_{ij}\hat E_{3y} = \kappa_{ij}\hat E_{ij},\ \kappa_{ij}=\pm1$. The second term in Eq.~\eqref{eq:mat_transform} acts within the two-dimensional subspaces $\{\hat E_{0x},\hat E_{3z}\}$ and $\{\hat E_{20},\hat E_{1y}\}$ which occurs only for $\kappa_{ij}=-1$. To compare to previous works, we employ the parametrization
\begin{align}
	\label{eq:qc_gf_isc}
	\hat g_\nu = \begin{bmatrix} g_0\hat\sigma_0+\bm g\cdot\hat{\bm{\sigma}} & [f_0\hat\sigma_0+\bm f\cdot\hat{\bm{\sigma}}]i\hat\sigma_y \\ [\bar f_0\hat\sigma_0+\bar{\bm {f}}\cdot\hat{\bm{\sigma}}^*]i\hat\sigma_y & \bar g_0\hat\sigma_0 + \bar{\bm{g}}\cdot\hat{\bm{\sigma}}^*\end{bmatrix}.
\end{align}
In general, the components $q$ and $\bar q$ are related by PHS, i.e. $\mathcal C \hat g_\nu(\hat{\bm{k}},\varepsilon) \mathcal C^{-1} = -\hat g_\nu(-\hat{\bm{k}},-\varepsilon)^*$ with $\mathcal C=[\hat\tau_1\otimes\hat\sigma_0]\mathcal K$, $\mathcal K$ denoting complex conjugation and $\hat{\bm{k}}$ the direction on the Fermi surface. In this case, we have $g_y,\bar g_y,f_z,\bar f_z=0$ as well as
\begin{alignat}{9}
	\bar g_0&=-&&g_0, \quad &&\bar g_x&&=&&g_x, \quad &&\bar g_z&&=-&&g_z, \notag\\
	\bar f_0&=&&f_0,  &&\bar f_x&&=-&&f_x, &&\bar f_y&&=&&f_y
\end{alignat}
and identify 
\begin{alignat}{4}
	4g_0 &=\operatorname{tr}(\hat g_\nu\hat E_{30}), \qquad && 4g_x  &&=\operatorname{tr}(\hat g_\nu\hat E_{0x}), \notag\\
	4g_z &=\operatorname{tr}(\hat g_\nu\hat E_{3z}), &&4f_0 &&=-i\operatorname{tr}(\hat g_\nu\hat E_{1y}), \notag\\
	\label{eqs:comps_gf_isc}
	4f_x &=i\operatorname{tr}(\hat g_\nu\hat E_{2z}), &&4f_y &&=-\operatorname{tr}(\hat g_\nu\hat E_{20}).
\end{alignat}
Note that $g_z,f_y$ are odd under a change of the valley index $\nu\to-\nu$. For numerical calculations we use a small Dynes broadening $\eta>0$ and introduce the shift $\varepsilon\to\varepsilon+i\eta$ removing exact degeneracies in the eigenvalues such that $\lambda_\pm^2\neq0, \lambda_+^2\neq\lambda_-^2$ are fulfilled in general. Physically, $\eta$ corresponds to a finite quasiparticle lifetime, e.g. due to elastic impurity scattering~\cite{PhysRevLett.41.1509}. 
\paragraph{Singlet pair correlation.} Since the rotation $\hat R_\nu$ acts within the subspace $\{\hat E_{20}, \hat E_{1y}\}$, i.e.
\begin{subequations}
	\begin{align}
		\hat R_\nu^\dag \hat E_{20}\hat R_\nu &= c_\alpha \hat E_{20} - \nu s_\alpha\hat E_{1y}, \\
		\hat R_\nu^\dag \hat E_{1y}\hat R_\nu &= c_\alpha \hat E_{1y} + \nu s_\alpha\hat E_{20},
	\end{align}
\end{subequations}
we first compute the coefficients in the primed basis
\begin{subequations}
	\begin{align}
		\frac{1}{4}\tr\,(\hat g_\nu'\hat E_{1y}) &= i\Delta c_\alpha\Big[u + v[3B^2+\varepsilon^2-\Delta^2]\Big],\\
		\frac{1}{4}\tr\,(\hat g_\nu'\hat E_{20}) &= -i\nu\Delta s_\alpha \Big[u + v[B^2+\varepsilon^2-\Delta^2]\Big]
	\end{align}
\end{subequations}
and find the component function $f_0$ in the unprimed basis with $\nu^2=1$
\begin{align}
	f_0 &=-i\Big[ - \frac{\nu s_\alpha}{4}\tr\,(\hat g_\nu'\hat E_{20}) + \frac{c_\alpha}{4}\tr\,(\hat g_\nu'\hat E_{1y})\Big] \notag\\
	\label{eq:f0_isc}
	&= \Delta \Big[u + v[B^2+\varepsilon^2-\Delta^2] + 2 c_\alpha^2 v B^2 \Big].
\end{align}
\paragraph{BCS limit.} The way we introduced $u,v$, they become singular as $\text{spec}(\hat A)$ becomes degenerate in the BCS limit. In order to take the limit correctly, we note that for $\lambda_+\to\lambda_-$, we have $s_+=s_-=1$ and $p(\lambda_j)=s_j$. Consequently using these identities before taking the limit, gives
\begin{subequations}
	\begin{align}
		v &= -\frac{1}{\lambda_+\lambda_-[\lambda_+ + \lambda_-]} \to -\frac{1}{2\lambda^3}, \\
		u &= \frac{\lambda_+\lambda_- + \lambda_+^2 + \lambda_-^2}{\lambda_+\lambda_-[\lambda_+ + \lambda_-]} \to \frac{3}{2\lambda}
	\end{align}
\end{subequations}
and Cayley-Hamiltonian $u + v\lambda^2 = 1/\lambda$ is satisfied. In the BCS limit $B_x,\beta,\alpha\to0$, we arrive at
\begin{align}
	f_0 &\to \Delta \left[ \frac{3}{2\lambda} - \frac{\lambda^2}{2\lambda^3}\right] = \frac{\Delta}{\lambda} = \frac{\Delta}{\sqrt{\varepsilon^2-\Delta^2}}.
\end{align}
In the same way, we find $g_0$
\begin{align}
	g_0 &= \varepsilon \Big[u + v[B^2+\varepsilon^2-\Delta^2] + 2vB^2\Big]
\end{align}
and the quasiclassical Green's function
\begin{align}
	\hat g_\nu \to \frac{\varepsilon\hat\tau_3\hat\sigma_0 + \Delta\hat\tau_1\hat\sigma_y}{\sqrt{\varepsilon^2-\Delta^2}}.
\end{align}
\paragraph{Connection to common notations.} To be consistent with common notations, we choose a different normalization via the mapping $\hat g_\nu\mapsto-i\pi N_0\hat\tau_3\hat g_\nu$ and find for $|\varepsilon|\leq\Delta$
\begin{align}
	\label{eq:wrong_gauge}
	\hat g_\nu = -\pi N_0 \frac{\varepsilon\hat\tau_0\hat\sigma_0 + i\Delta\hat\tau_2\hat\sigma_y}{\sqrt{\Delta^2-\varepsilon^2}}.
\end{align}
The standard BCS Green's function follows from a global $U(1)$ gauge rotation in Nambu space
\begin{align}
	\hat g_\nu\mapsto\hat U\hat g_\nu \hat U^\dag\quad\text{with}\quad\hat U=\exp(i\varphi \hat\tau_3/2),
\end{align}
under which the superconducting order parameter transforms as $\Delta\mapsto \Delta e^{i\varphi}$. With $\varphi=\pi/2$, Eq.~\eqref{eq:wrong_gauge} becomes
\begin{align}
	\label{eq:bcs_mybasis}
	\hat g_\nu = -\pi N_0 \frac{\varepsilon\hat\tau_0\hat\sigma_0 - \Delta\hat\tau_2\hat\sigma_y}{\sqrt{\Delta^2-\varepsilon^2}}.
\end{align}
It is connected to the Green's function in the basis $\phi = [c_{\bm{k},\uparrow}, c_{-\bm{k},\downarrow}^\dag, c_{\bm{k},\downarrow}, -c^\dag_{-\bm{k},\uparrow}] = \hat Q \psi_{\rm S}$ with $\hat Q = \hat S \hat P$, where $P_{ij}=\delta_{i,\pi(j)}$ follows from $\pi=[1,4,2,3]$ and $\hat S=\text{diag}(\{1,1,1,-1\})$. We recover the expected representation, $-\hat Q \hat\tau_2\hat\sigma_y \hat Q^T\to\hat\sigma_0\hat\tau_1$, in the basis $\phi$, compare Ref.~\cite{PhysRevB.103.155407}.

\begin{figure}[ht]
	\centering
	\includegraphics[scale=1]{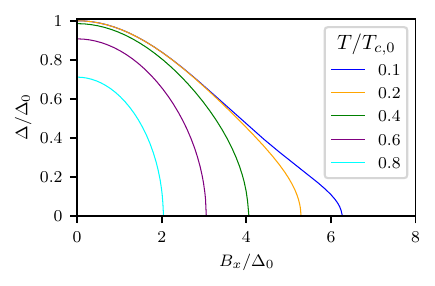}
	\caption{Superconducting gap as a function of in-plane magnetic field strength for temperatures as indicated.}
	\label{fig:selfconsistency}
\end{figure}
\section{Self-consistency equation}
\label{sec:selfconsistency}
We compute the superconducting gap self-consistently according to
\begin{align}
    \Delta = \lambda \sum_{\bm k}\langle c_{-\bm k,\downarrow} c_{\bm k,\uparrow}\rangle,
\end{align}
with $\lambda>0$ the dimensionless strength of a $\bm k$-independent attractive pairing interaction. In the same approximation as in the main text, the self-consistency equation reads
\begin{align}
    \frac{\Delta}{\lambda} = -2\pi i T \sum_{n=0}^{N_c} f_0(i\omega_n),
\end{align}
with $\omega_n=(2n+1)\pi T$ the fermionic Matsubara frequencies, $k_{\rm B}=1$, $N_c$ a large energy cut-off and $f_0$ the correlation function from Eq.~\eqref{eq:f0_isc}.
For faster numerical convergence and independence of a specific model for the cut-off frequency, we subtract $2\pi T\sum_n \Delta/\omega_n$ from both sides and identify the definition of the critical temperature at zero field $T_{c,0}$ in the limit of $\Delta,B_x\to0$, in which $f_0\to i\Delta/\omega_n$ on the right side. With $N_c(T)=\lfloor \omega_c/(2\pi T)-1/2\rfloor\approx\omega_c/(2\pi T)$, we find
\begin{align}
	\frac{1}{\lambda} - \sum_{n=0}^{N_c(T)} \frac{2\pi T}{\omega_n} &= \sum_{n=0}^{N_c(T_{c,0})} \frac{1}{n+\frac{1}{2}} - \sum_{n=0}^{N_c(T)} \frac{1}{n+\frac{1}{2}} \notag\\
	&= \psi\left(\frac{\omega_c}{2\pi T_{c,0}} + \frac{3}{2}\right) - \psi\left( \frac{\omega_c}{2\pi T} + \frac{3}{2}\right)\notag\\
	&\overset{\omega_c\to\infty}{\longrightarrow} \log(\frac{T}{T_{c,0}}).
\end{align}
In terms of the renormalization scale $T_{c,0}$, the self-consistency equation reads
\begin{align}
	\log(\frac{T}{T_{c,0}}) = \frac{2\pi T}{\Delta}\sum_n \left[\Im(f_0(i\omega_n)) - \frac{\Delta}{\omega_n}\right]
\end{align}
which we solve numerically. In Fig.~\ref{fig:selfconsistency}, we show the solution of the self-consistency equation for some chosen values of $T/T_{c,0}$.

\section{Symmetry allowed hopping elements}
\label{sec:symmetry}
We determine the symmetry-allowed structure of the hopping matrix $\hat V(\nu\bm k) = v_0^\nu\hat\sigma_0 + \bm v^\nu\cdot\hat{\bm{\sigma}}$ by systematically combining the symmetry constraints of the substrate and the impurity. The key object is the \emph{common unbroken symmetry group} $\tilde H_{\rm eff}$ that remains unbroken in the coupled system. This means that we look for the hopping matrix elements fulfilling
\begin{align}
	R_{\rm S}(g) \hat V(\bm k) R_{\rm imp}(g)^{-1} = \hat V(g\bm k)
\end{align}
where $R_i(g)$ denotes the representation of the symmetry element $g\in\tilde H_{\rm eff}$ in subsystem $i=\{{\rm S, imp}\}$. The effective symmetry group is defined by
\begin{align}
	\tilde H_{\rm sub} = \tilde H_{\rm site}\cap\tilde G_{\bm k}, \qquad \tilde H_{\rm eff} = \tilde H_{\rm sub}\cap\tilde H_{\rm imp},
\end{align}
where $G_{\bm k}$ is the little group at momentum $\bm k$, $H_{\rm site}$ the point group symmetry of the site at which the impurity is located in the substrate and $H_{\rm sub}, H_{\rm imp}$ the symmetry groups of substrate and impurity. We denote the double group lift of $X_i$ by $\tilde X_i$.

\paragraph{Substrate symmetries and valley projection.} Let $g=\{R\mid \tau\}$ be an element of the space group, with unitary representation $\hat U_{\rm S}(g)$ acting on the single-particle Hilbert space. The substrate Hamiltonian $\hat h_0(\bm k)$ satisfies
\begin{align}
	\hat U_{\rm S}(g)\,\hat h_0(\bm k)\,\hat U_{\rm S}(g)^{-1} = \hat h_0(g\bm k), \quad g\in \tilde G_{\bm k}.
\end{align}
In a Bloch basis $\ket{\alpha,\bm k,\sigma}$, the symmetry action is encoded by the sewing matrix
\begin{align}
	\hat U_{\rm S}(g) \ket{\alpha,\bm k,\sigma} = \sum_{\beta,\sigma'} D_{\beta\sigma',\alpha\sigma}(g,\bm k) \ket{\beta,g\bm k,\sigma'}.
\end{align}
Due to spin-orbit coupling, orbital and spin degrees of freedom do not factor globally in general. However, after projection onto a low-energy subspace corresponding to a single valence-band orbital at a given valley $\nu=\pm1$, the orbital part reduces to a one-dimensional character $\chi_\nu(g;\bm k)\in U(1)$. On the projected subspace one obtains
\begin{align}
	\label{eq:1drep}
	\hat U_{\rm S}(g)\big|_\nu = \chi_\nu(g;\bm k) D^{(1/2)}(R),
\end{align}
where $D^{(1/2)}(R)$ is the spin-$\tfrac12$ representation of the rotation $R$. For the $K$ and $K'$ valleys of a 1H transition-metal dichalcogenide, the spin-less little group is $C_{3h}=\langle C_3^n, \sigma_h\rangle, n=0,1,2$. According to Refs.~\cite{PhysRevB.88.085433,PhysRevLett.108.196802}, the valence band edge at $\nu\bm K$ is described to good accuracy by the wave function
\begin{align}
	\ket{\phi^\nu} = \frac{1}{\sqrt2} \Big[\ket{d_{x^2-y^2}} + i\nu \ket{d_{xy}}\Big].
\end{align}
This state transforms according to a one-dimensional orbital representation and symmetries are represented as in Eq.\eqref{eq:1drep} in the basis
\begin{align}
	\mathcal B_\nu=\{\ket{\phi^\nu,\uparrow}, \ket{\phi^\nu,\downarrow}\}.
\end{align}
At the hollow site of the 1H lattice, the horizontal mirror $\sigma_h$ acts trivially on the orbital part of $\ket{\phi^\nu}$ since both $d_{x^2-y^2}$ and $d_{xy}$ are even under $\sigma_h$. In contrast, $C_3$ contributes a valley-dependent phase associated with $m_\nu=2\nu$, the effective $C_3$ angular-momentum quantum number of the chiral valence-band state $\ket{\phi^\nu}$. More explicitly, the spin-sector representations are
\begin{subequations}
	\begin{align}
		D^{(1/2)}(C_3^n) &= \exp\left(-i\frac{n\pi}{3}\sigma_z\right), \\
		D^{(1/2)}(\sigma_h) &\sim i\sigma_z,
	\end{align}
\end{subequations}
where $\sim$ denotes equality up to an overall phase. For a local orbital centered at $\bm r_0$ and for a valley momentum $\bm K$ satisfying $g\bm K = \bm K + \bm G$, the orbital character obeys
\begin{align}
	\chi_\nu(g;\bm K) = e^{-i\bm G\cdot\bm r_0} \eta_\nu(g),
\end{align}
with $\eta_\nu(g)$ the \textit{onsite} character under the point-group part of $g$ and $\exp(-i\bm G\cdot\bm r_0)$ a phase arising from lattice translations. For a $C_3$ rotation,
\begin{align}
	\chi_\nu(C_3^n) = \exp\left(-i\Big[\bm G_n\cdot\bm r_0 + m_\nu\frac{2\pi n}{3}\Big]\right).
\end{align}
Since the orbital sector contributes only as a scalar phase factor in the projected representation, we can classify the symmetry-allowed hopping matrix elements using only the spin structure.

\paragraph{Classical impurity.} We choose a spin basis in which the impurity moment $\bm J\parallel\hat z$. The impurity is invariant only under transformations that preserve this direction. In particular, continuous spin rotations about the $z$-axis form
\begin{align}
	\tilde H_{\rm imp} = \{ e^{-i\theta\sigma_z/2} \,|\, \theta\in[0,2\pi) \} \cong U(1).
\end{align}
The fixed classical moment explicitly breaks time-reversal symmetry.

\paragraph{Results for the hopping matrix at a valley fixed point.} At a fixed point $g\bm K=\bm K$, the symmetry constraint reduces to
\begin{align}
	R_{\rm S}(g) \hat V(\nu\bm K) R_{\rm imp}(g)^{-1} = \hat V(\nu\bm K).
\end{align}
If $C_3 \in\tilde H_{\rm eff}$, both substrate and impurity representations share the same rotation axis generated by $\sigma_z$. Invariance under this generator implies $[\hat V(\nu \bm K),\sigma_z]=0$. In our case, the in-plane field $B_x\neq0$ breaks all non-trivial common symmetries and $\tilde H_{\rm eff}=E$, leading to
\begin{align}
	\hat V(\nu\bm K) = v_0^\nu\sigma_0 + \bm v^\nu\cdot\bm\sigma \in\mathbb C^{2\times2}.
\end{align}

\paragraph{Nambu extension.} In the basis from the main text $\psi_{\rm S}(\bm k) = \big[c_{\bm k\uparrow},c_{\bm k\downarrow},c^\dag_{-\bm k\uparrow},c^\dag_{-\bm k\downarrow}\big]^T$, particle-number-conserving tunneling lifts
\begin{align}
	\hat V(\bm k) \longrightarrow \begin{bmatrix} \hat V(\bm k) & \mathbb 0 \\ \mathbb 0 & -\hat V^*(-\bm k)\end{bmatrix},
\end{align}
where the relative minus sign reflects the particle-hole structure of Nambu space.

\section{Analysis of the YSR Green's function}
\label{sec:ysr_analysis}
In general the matrix representation of the YSR Greens function $\hat G\equiv\hat G_{\rm YSR}$, the bare impurity Green's function $\hat g\equiv\hat g_{\rm imp}=\text{diag}\,(\hat g^{\rm ee},\hat g^{\rm hh})$, the self energy $\hat\Sigma$ and any product of them can be decomposed into
\begin{align}
	\label{eq:blockparam}
	\hat X = \begin{bmatrix} \hat X^{\rm ee} & \hat X^{\rm eh} \\ \hat X^{\rm he} & \hat X^{\rm hh}\end{bmatrix}, \quad \hat X^{\tau\tau'} = X^{\tau\tau'}_0\hat\sigma_0 + \bm X^{\tau\tau'}\cdot\hat{\bm{\sigma}}
\end{align}
with $X=\{G,\Sigma,g,...\}$. We employ the notation with supercripts $\tau=1$(e) for electrons and $\tau=-1$(h) for holes as well as $\bar\tau=-\tau$. The self energy due to coupling to the ISC is thus decomposed into
\begin{subequations}
	\begin{align}
		\hat\Sigma^{\rm ee} &= \phantom{-}\frac12\sum_\nu \hat V(\nu\bm K)\hat g_\nu^{\rm ee}\hat V^\dag(\nu\bm K), \\
		\hat\Sigma^{\rm eh} &= -\frac12\sum_\nu \hat V(\nu\bm K)\hat g_\nu^{\rm eh} \hat V^T(\bar\nu\bm K), \\
		\hat\Sigma^{\rm he} &= -\frac12\sum_\nu \hat V^*(\bar\nu\bm K)\hat g_\nu^{\rm he} \hat V^\dag(\nu\bm K), \\
		\hat\Sigma^{\rm hh} &= \phantom{-}\frac12\sum_\nu \hat V^*(\bar\nu\bm K)\hat g_\nu^{\rm hh}\hat V^T(\bar\nu\bm K),
	\end{align}
\end{subequations}
where $\bar\nu=-\nu$ and
\begin{subequations}
	\label{eqs:gprime}
	\begin{align}
		\hat g_\nu^{\tau\tau} &= \pi N_0 \big[ g_0\hat\sigma_0 + \tau g_x \hat\sigma_x + g_z\hat\sigma_z\big], \\
		\hat g_\nu^{\tau\bar\tau} &= \pi N_0 \big[-if_y\hat\sigma_0 + \tau f_0 i\hat\sigma_y - if_x \hat\sigma_z\big]
	\end{align}
\end{subequations}
according to Eqs.~\eqref{eqs:comps_gf_isc} with an absorbed factor of $-i$.

\subsection{Bound state energies and effective spin-splitting}
\label{sec:bound_states_app}
Bound states follow from $\det\,(\hat g^{-1} - \hat\Sigma)=0$. Under condition of $\det\,([\hat G^{-1}]^{\bar\tau\bar\tau})\neq0$, a Schur complement reduction yields the reduced equation for bound states
\begin{align}
	\label{eq:reduced_cond}
	0 = \det\,([\hat G^{\tau\tau}]^{-1})
\end{align}
with the Schur-complement
\begin{align}
	\label{eq:effective_ee}
	[\hat G^{\tau\tau}]^{-1} &= [\hat g^{\rm\tau\tau}]^{-1} - \hat\Sigma^{\tau\tau} - \hat\Omega^{\tau\tau}
\end{align}
and
\begin{align}
	\label{eq:Omega_Lambda}
	\hat\Omega^{\tau\tau}\equiv\hat\Sigma^{\tau\bar\tau}\hat\Lambda^{\bar\tau\bar\tau}\hat\Sigma^{\bar\tau\tau}, \quad \hat \Lambda^{\tau\tau}\equiv [[\hat g^{\tau\tau}]^{-1} - \hat\Sigma^{\tau\tau}]^{-1}.
\end{align}
In Eq.~\eqref{eq:effective_ee}, we identify two things. Firstly, $[\hat G^{\tau\tau}]^{-1}$ describes an effective/dressed inverse Green's function of the $\tau$-block and secondly, there exists a substrate induced energy dependent spin-splitting $\bm B_\tau\equiv\bm\Sigma^{\tau\tau}+\bm\Omega^{\tau\tau}\in\mathbb C^3$ in the $\tau$-block. In general, it reads
\begin{align}
	\bm B_\tau = \bm\Sigma^{\tau\tau} &+ \bm\Lambda^{\bar\tau\bar\tau} \left[ \Sigma_0^{\tau\bar\tau}\Sigma_0^{\bar\tau\tau} - \bm\Sigma^{\tau\bar\tau}\cdot\bm\Sigma^{\bar\tau\tau} \right] \notag\\
	& + \bm\Sigma^{\tau\bar\tau}\left[ \Sigma_0^{\bar\tau\tau} \Lambda_0^{\bar\tau\bar\tau} + \bm\Sigma^{\bar\tau\tau} \cdot \bm\Lambda^{\bar\tau\bar\tau}\right]\notag\\
	& + \bm\Sigma^{\bar\tau\tau}\left[ \Sigma_0^{\tau\bar\tau} \Lambda_0^{\bar\tau\bar\tau} + \bm\Sigma^{\tau\bar\tau} \cdot \bm\Lambda^{\bar\tau\bar\tau}\right]\notag\\
	\label{eq:Ba}
	& + i \Big\{\Sigma^{\tau\bar\tau}_0 \left[ \bm\Lambda^{\bar\tau\bar\tau} \times \bm\Sigma^{\bar\tau\tau}\right] + \Sigma^{\bar\tau\tau}_0 \left[ \bm\Sigma^{\tau\bar\tau}\times\bm\Lambda^{\bar\tau\bar\tau}\right]\notag\\
	&\phantom{+ i \Big\{} +  \Lambda_0^{\bar\tau\bar\tau} \left[ \bm\Sigma^{\tau\bar\tau}\times\bm\Sigma^{\bar\tau\tau}\right] \Big\}.
\end{align}
The Schur-complement then reads
\begin{align}
	[\hat G^{\tau\tau}]^{-1}(\varepsilon) &= E_\tau(\varepsilon)\hat\sigma_0 - \big[\tau\bm B_{\rm imp} + \bm B_\tau(\varepsilon)\big]\cdot\hat{\bm{\sigma}},
\end{align}
with the scalar term $E_\tau\equiv [\hat g^{-1}]^{\tau\tau}_0 - \Sigma^{\tau\tau}_0 - \Omega^{\tau\tau}_0\in\mathbb C$ explicitly given by
\begin{align}
	E_\tau = &[\hat g^{-1}]^{\tau\tau}_0 - \Sigma^{\tau\tau}_0 - \Sigma^{\tau\bar\tau}_0 \Lambda^{\bar\tau\bar\tau}_0 \Sigma^{\bar\tau\tau}_0 \notag\\
	& - \bm\Lambda^{\bar\tau\bar\tau}\cdot \big[ \Sigma^{\tau\bar\tau}_0 \bm\Sigma^{\bar\tau\tau} + \Sigma^{\bar\tau\tau}_0 \bm\Sigma^{\tau\bar\tau}\big] \notag\\
	& -\Lambda^{\bar\tau\bar\tau}_0 \bm\Sigma^{\tau\bar\tau}\cdot\bm\Sigma^{\bar\tau\tau} - i \big[ \bm\Sigma^{\tau\bar\tau}\times\bm\Lambda^{\bar\tau\bar\tau}\big] \cdot \bm\Sigma^{\bar\tau\tau}.
\end{align}
In the subgap regime there exist true bound states with poles on the real axis, i.e. $\varepsilon_0\in\mathbb R$ for $\eta\to0$. In this case, the bound state condition from Eq.~\eqref{eq:reduced_cond} reduces to
\begin{align}
	\label{eq:boundstates_real}
	\mu_\tau(\varepsilon) = \pm |\bm J^\tau_{\rm eff}(\varepsilon)|
\end{align}
with $\mu_\tau=\Re(E_\tau)$ the effective potential and $\bm J^\tau_{\rm eff}\equiv \Re(\tau\bm B_{\rm imp}+\bm B_\tau(\varepsilon))$ the effective magnetic moment. We define these quantities using the real part to extend the study of their behaviour to the entire energy range. The complement block of the inverse Green's function reads
\begin{align}
	[\hat G^{-1}]^{\bar\tau\bar\tau}&= [\varepsilon -\bar\tau U_0 - \Sigma^{\bar\tau\bar\tau}_0]\hat\sigma_0 + [\bar\tau\bm B_{\rm imp} - \bm\Sigma^{\bar\tau\bar\tau}]\cdot\hat{\bm{\sigma}}
\end{align}
and invertibility requires
\begin{align}
	\label{eq:cond_invertibility_hh}
	\varepsilon - \Sigma^{\bar\tau\bar\tau}_0 - \bar\tau U_0 \neq \pm|\bar\tau\bm B_{\rm imp} - \bm\Sigma^{\bar\tau\bar\tau}|.
\end{align}
In Fig.~\ref{fig:effective_spinsplitting_ISC}, we plot the effective magnetic moment of an impurity on an ISC for some chosen parameters.
\begin{figure*}[ht]
	\centering
	\includegraphics[scale=1]{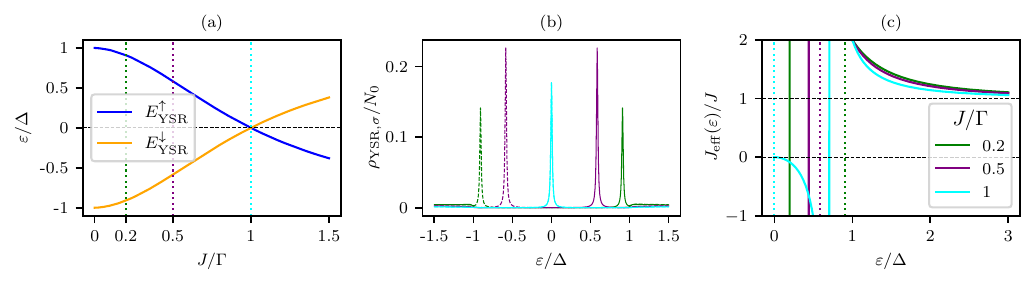}
	\caption{(a) Spin-resolved YSR bound state energies. (b) Spin-resolved local density of states shown by solid (dashed) lines for $\sigma=\uparrow (\downarrow)$. The color code indicates the values of $J/\Gamma$ as in the other plots. (c) Effective magnetic moment for indicated values of $J/\Gamma$ (solid) and corresponding bound state energy (dotted). In (b) and for $\varepsilon\geq\Delta$ in (c), we introduce $\varepsilon\to\varepsilon+i\eta$ with $\eta=10^{-2}\Delta$. All plots are generated using $\Gamma=50\Delta$ and $U_0=0$.}
	\label{fig:effective_spinsplitting_BCS}
\end{figure*}
\paragraph{Example: BCS YSR states.}
\label{par:bcs_example}
 Following Ref.~\cite{PhysRevB.103.155407}, in the basis $\bm{\tilde c}_{\bm k}=[c_{\bm k,\uparrow}, c_{-\bm k,\downarrow}^\dag, c_{\bm k,\downarrow}, -c_{-\bm k,\uparrow}^\dag]^T$ we have
\begin{align}
	\hat g_{\rm BCS} &= \frac{-\pi N_0}{\sqrt{\Delta^2-\varepsilon^2}}\hat\sigma_0\big[\varepsilon\hat\tau_0+\Delta\hat\tau_1 \big], \\
	\hat g_{\rm imp}^{-1} &= \varepsilon\hat\sigma_0\hat\tau_0 - U_0 \hat\sigma_0\hat\tau_3 - J\hat\sigma_3\hat\tau_0
\end{align}
with $\bm J$ projected along the impurity quantization axis and $|\varepsilon|\leq\Delta$. For spin-independent hopping, $\hat V=t \hat\sigma_0\hat\tau_3$ with $\Gamma\equiv \pi N_0 t^2$, the self-energy reads
\begin{align}
	\hat\Sigma = -\frac{\Gamma\varepsilon}{\sqrt{\Delta^2-\varepsilon^2}}\hat\sigma_0\hat\tau_0 + \frac{\Gamma\Delta}{\sqrt{\Delta^2-\varepsilon^2}} \hat\sigma_0\hat\tau_1.
\end{align}
It contains no spin dependent terms, $\bm\Sigma^{\tau\tau}=\bm 0$ and the dynamic spin-splitting solely stems from pairing correlations, i.e. $\bm B_\tau=\bm\Omega_\tau$. Following Eq.~\eqref{eq:Omega_Lambda}, we have
\begin{align}
	\hat\Lambda^{\bar\tau\bar\tau} &= \left[a_{\bar\tau}\hat\sigma_0 - J\hat\sigma_3\right]^{-1},\\
	\hat\Omega^{\tau\tau} &= \frac{\Gamma^2\Delta^2}{\Delta^2-\varepsilon^2}\frac{a_{\bar\tau}\hat\sigma_0 + J\hat\sigma_3}{a_{\bar\tau}^2-J^2}
\end{align}
with $a_\tau\equiv\varepsilon [1+ \Gamma/\sqrt{\Delta^2-\varepsilon^2}] - \tau U_0$. Note that $a_{\bar\tau}(-\varepsilon)=-a_\tau(\varepsilon)$. In the basis $\tilde{\bm{d}}_\tau=[d_{\sigma_\tau}, \tau d^\dag_{\sigma_{\bar\tau}}]^T$ with $\sigma_\tau=\uparrow,\downarrow$ for $\tau={\rm e(+),h(-)}$, we have
\begin{align}
	[\hat G^{\tau\tau}]^{-1}(\varepsilon) = &\left[a_\tau - \frac{\Gamma^2\Delta^2}{\Delta^2-\varepsilon^2}\frac{a_{\bar\tau}}{a_{\bar\tau}^2-J^2}\right]\hat\sigma_0 \notag\\
	\label{eq:bcs_schur}
	 - &J\left[1 + \frac{\Gamma^2\Delta^2}{\Delta^2-\varepsilon^2}\frac{1}{a_{\bar\tau}^2-J^2} \right]\hat\sigma_3.
\end{align}
Note that in this basis, PHS manifests itself by $\bm J^{\rm e}_{\rm eff}(\varepsilon) = \bm J^{\rm h}_{\rm eff}(-\varepsilon)$ and $\mu_{\rm e}(\varepsilon)=-\mu_{\rm h}(-\varepsilon)$. Due to spin-rotation invariance of the BCS superconductor, YSR bound states appear pairwise and the solutions for $\sigma=\uparrow,\downarrow$ are linked to opposite energies with $\sigma=\uparrow,\downarrow$ corresponding to the $\pm$ solutions following from Eq.~\eqref{eq:bcs_schur}. One can see explicitly, that both $\tau$-blocks produce the same solution by noting that
\begin{align}
	&[\hat G^{\bar\tau\bar\tau}]^{-1}(-\varepsilon) = -\hat\sigma_1[\hat G^{\tau\tau}]^{-1}(\varepsilon)\hat\sigma_1 \notag\\
	&\Rightarrow\det\,([\hat G^{\bar\tau\bar\tau}]^{-1}(-\varepsilon)) = \det\,([\hat G^{\tau\tau}]^{-1}(\varepsilon)).
\end{align}
Thus, it suffices to use only one $\tau$-block. In the sub-gap regime, bound states follow from Eq.~\eqref{eq:bcs_schur} by
\begin{align}
	\label{eq:cond_bs_BCS}
	\tr\,([\hat G^{\tau\tau}]^{-1}) = \pm\tr\,([\hat G^{\tau\tau}]^{-1}\hat\sigma_3)
\end{align}
under the condition $\det\,([\hat G^{-1}]^{\bar\tau\bar\tau})\neq0\Leftrightarrow a_{\bar\tau}^2-J^2\neq0$ which is equivalent to Eq.~\eqref{eq:boundstates_real} with Eq.~\eqref{eq:cond_invertibility_hh}. More explicitly, for $U_0=0 \Rightarrow a_{\rm e}=a_{\rm h}=a$ and $\tau={\rm e}$, bound states follow from $a - J_\sigma = \pm\Gamma\Delta/\sqrt{\Delta^2-\varepsilon^2}$ with $J_\uparrow=-J_\downarrow=J$. Reinserting $a$, we find
\begin{align}
	\label{eq:bs_cond}
	& 2\Gamma\varepsilon \big[\varepsilon - J_\sigma\big] + \big[[\varepsilon - J_\sigma]^2-\Gamma^2\big] \sqrt{\Delta^2-\varepsilon^2} = 0
\end{align}
which is equivalent to the poles of Eq. (9) in Ref.~\cite{PhysRevB.103.155407} for $U_0=0$. Using this approach, we observe a renormalized frequency dependent spin-splitting
\begin{align}
	\label{eq:J_eff}
	\frac{J^{\rm e}_{\rm eff}(\varepsilon)}{J}\equiv \Re\!\left( 1 + \frac{\Gamma^2\Delta^2}{\Delta^2-\varepsilon^2}\frac{1}{a^2-J^2}\right)
\end{align}
visualized together with the bound state energies in Fig.~\ref{fig:effective_spinsplitting_BCS} for $\varepsilon>0$ since $J^{\rm e}_{\rm eff}(\varepsilon)=J^{\rm e}_{\rm eff}(-\varepsilon)$ for $U_0=0$. The effective field vanishes for $J/\Gamma=1$ at the bound state energy $\varepsilon_*=0$. Outside the sub-gap regime, for $\varepsilon\gg\Delta$, the influence of the pair correlation on the effective magnetic moment decays until $J_{\rm eff}(\varepsilon)\to J$.

\subsection{Effective spin-splitting in an ISC}
\label{sec:effective_field}
\begin{figure*}[ht]
	\centering
	\includegraphics[scale=1]{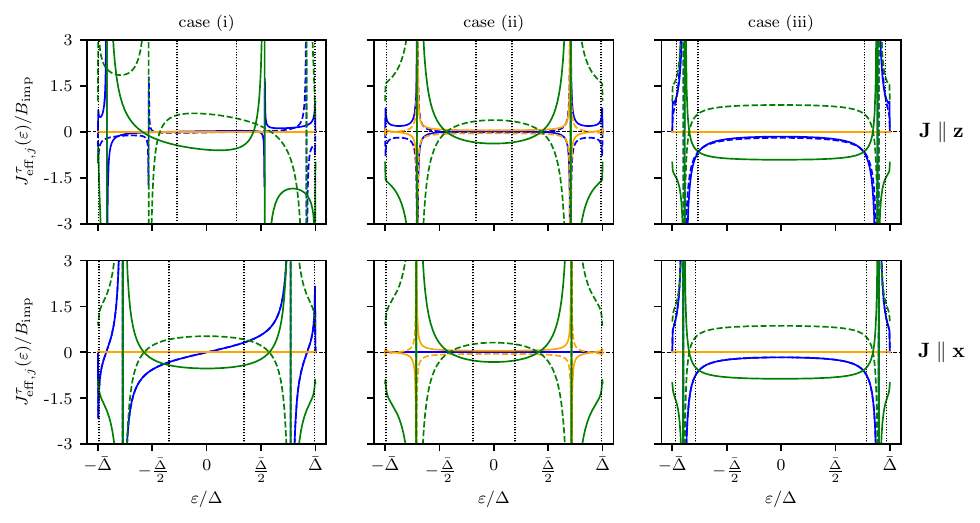}
	\caption{Components $j=x,y,z$ (blue, orange, green) of the effective magnetic moment projected onto the impurity spin at $J/\Gamma=1$ for $\tau={\rm e,h}$ (solid, dashed) and $\theta=0,\pi/2$ (top, bottom). Indicated with black dotted lines are the corresponding bound state energies. The plots are generated with $\{B_x,\beta,\pi N_0|v_1|^2,U_0\}=\{3,7,50,
	0\}\Delta$, $|v_1|^2/|v_2|^2=10$ and $\eta=10^{-30}\Delta$.}
	\label{fig:effective_spinsplitting_ISC}
\end{figure*}

We compute the $\tau\tau'$-blocks of the self-energies projected onto the impurity spin. Note that we use the component functions from Eq.~\eqref{eqs:gprime}. We define the tunneling rates $\Gamma
\equiv\pi N_0[v_1^2+v_2^2]$, $\Gamma_-\equiv\pi N_0[v_1^2-v_2^2]$, $\Gamma_\times\equiv\pi N_0 v_1 v_2$. Different to the BCS case, we have $\bm\Sigma^{\tau\tau}\neq\bm 0$ in general. More explicitly, according to Eq.~\eqref{eq:hoppingmatrix}, we have
\begin{subequations}
	\begin{align}
		\hat\Sigma^{\tau\tau}_{\rm (i)} = &\Gamma g_0 \hat\sigma_0 + [\tau\Gamma_- g_x c_\vartheta - 2\Gamma_\times g_0 s_\vartheta]\hat\sigma_x \notag\\
		&+ [\tau\Gamma_- g_x s_\vartheta + 2\Gamma_\times g_0 c_\vartheta]\hat\sigma_z, \\
		\hat\Sigma^{\tau\tau}_{\rm (ii)} = &\Gamma g_0 \hat\sigma_0 \notag\\
		&+ \tau  g_x [\Gamma_- c_\vartheta\hat\sigma_x + 2\Gamma_\times\hat\sigma_y + \Gamma_-s_\vartheta\hat\sigma_z], \\
		2\hat\Sigma^{\tau\tau}_{\rm (iii)} = &\Gamma g_0\hat\sigma_0 + \big[\tau\Gamma g_x c_\vartheta - \Gamma_- g_z s_\vartheta\big] \hat\sigma_x \notag\\
		&+ \big[\tau\Gamma g_x s_\vartheta + \Gamma_- g_z c_\vartheta \big]\hat\sigma_z
	\end{align}
\end{subequations}
as well as
\begin{subequations}
	\begin{align}
		\hat\Sigma^{\tau\bar\tau}_{\rm (i)} =& 2i\Gamma_\times f_x \hat\sigma_0 - \tau\Gamma_-  f_0 i\hat\sigma_y \notag\\
		& - i\Gamma f_x [s_\vartheta\hat\sigma_x - c_\vartheta\hat\sigma_z], \\
		\hat\Sigma^{\tau\bar\tau}_{\rm (ii)} =& -2\tau\Gamma_\times f_x\hat\sigma_0 + \tau\Gamma f_0 i\hat\sigma_y \notag\\
		&+ i\Gamma_- f_x [s_\vartheta\hat\sigma_x - c_\vartheta\hat\sigma_z], \\
		2\hat\Sigma^{\tau\bar\tau}_{\rm (iii)} =& i\Gamma_- f_y\hat\sigma_0 - \tau\Gamma f_0 i\hat\sigma_y \notag \\
		&- i\Gamma f_x\big[s_\vartheta\hat\sigma_x - c_\vartheta\hat\sigma_z\big].
	\end{align}
\end{subequations}
Furthermore, we have
\begin{align}
	[\hat\Lambda^{\tau\tau}_j]^{-1} &= [\varepsilon-\tau U_0 - \Sigma^{\tau\tau}_{0,j}]\hat\sigma_0 - [\tau \bm B_{\rm imp} + \bm\Sigma^{\tau\tau}_j]\cdot\hat{\bm{\sigma}}.
\end{align}
Different to the BCS case, the ISC breaks spin-rotation invariance and the effective magnetic moment acquires transverse components which generically lifts the spin degeneracy. We observe four YSR states, pair-wise linked to opposite energies by PHS as shown for some parameters in Fig.~\ref{fig:effective_spinsplitting_ISC}. 

\paragraph{PHS symmetry representation.} In the basis $\psi_{\rm imp}=[d_\uparrow,d_\downarrow,d^\dag_\uparrow,d^\dag_\downarrow]$ from the main text, PHS requires $\bm J^{\rm e}_{\rm eff}(\varepsilon_0)=-\bm J^{\rm h}_{\rm eff}(-\varepsilon_0)$ at the bound state energy. In the following, we consider the function over the entire energy range. We observe that PHS does not always factor in the same way and we introduce $\mathcal C = \mathcal S \mathcal P$ with
\begin{subequations}
	\label{eqs:phs_comp}
	\begin{alignat}{3}
		&\mathcal P \bm J^\tau_{\rm eff}(\varepsilon) &&= \sigma_P \bm J^\tau_{\rm eff}(-\varepsilon) 
		\intertext{as well as}
		&\mathcal S \bm J^\tau_{\rm eff}(\varepsilon) &&= \sigma_S \bm J^{\bar\tau}_{\rm eff}(\varepsilon)
	\end{alignat}
\end{subequations}
where $\sigma_S,\sigma_P\in\{\pm1\}$ component-wise. In Eqs.~\eqref{eqs:phs_comp}, $\sigma_S,\sigma_P$ are signature matrices in three dimensional component space. \emph{Note: Here, we interpret $\mathcal C,\mathcal S,\mathcal P$ as transformation rules acting on the coefficient functions $\mu_\tau(\varepsilon),\bm J^\tau_{\rm eff}(\varepsilon)$ rather than as operators in Hilbert space.} Since $\mathcal S^2 = \mathcal P^2= 1$, they generate $\mathbb Z_2$ groups and $\sigma_S,\sigma_P$ correspond to their one-dimensional characters. PHS requires $\sigma_C=-1$ for each component. It is worth noting that the factorization of $\mathcal C$ does not imply that $\mathcal S$ and $\mathcal P$ act trivially in component space, as illustrated in Fig.~\ref{fig:effective_spinsplitting_ISC}. In the absence of valley-symmetry, we do not observe a PHS relation of the form given in Eqs.~\eqref{eqs:phs_comp}. In $\bm\Sigma^{\tau\tau}_{\rm (iii)},\bm\Sigma^{\tau\bar\tau}_{\rm (iii)}$, we find that the component functions $g_z,f_y$ respectively appear with $\sigma_S=+$ as well as
\begin{subequations}
	\begin{align}
		\Re(g_z)\colon \sigma_P=+, \qquad \Im(g_z)\colon \sigma_P=-, \\
		\Re(f_y)\colon \sigma_P=-, \qquad \Im(f_y)\colon \sigma_P=+.
	\end{align}
\end{subequations}
In the effective magnetic moment $\bm J^\tau_{\rm eff}$, more explicitly in the term $\Re(\bm\Sigma^{\tau\tau}_{\rm (iii)})$, this leads to $\sigma_C=+$ in the $x$-component $\sim\Re(g_z)$ for $\theta=\pi/2$, thus obscuring the simple form of the PHS relation. This does not imply a breaking of PHS, but rather that PHS is no longer represented in a straight forward component-wise form anymore. More explicitly, bound states still appear in pairs $\{E,-E\}$.

\subsection{Bound state crossings} 
\label{sec:bs_symmetry}
In the subgap regime and for $\eta\to0$, a bound state at $\varepsilon_0\in\mathbb R$ in sector $\tau$ follows from $\lambda_\tau^{s}(\varepsilon_0)=0$ with
\begin{align}
    \lambda_\tau^{s}(\varepsilon) = \mu_\tau(\varepsilon) - s |\bm J^\tau_{\rm eff}(\varepsilon)| \qquad s\in\{\pm1\}.
\end{align}
An exact crossing at the bound state energy means that there exist $s,t\in\{\pm1\}$ such that $\lambda_{\rm e}^{s}(\varepsilon_0)=0$ and $\lambda_{\rm h}^{t}(\varepsilon_0)=0$ stating that the two sectors support bound states at the same energy. If, in addition the corresponding eigenvectors coincide and $\hat h^{\rm ee}(\varepsilon_0)$ and $\hat h^{\rm hh}(\varepsilon_0)$ are simultaneously diagonalizable, i.e.
\begin{align}
	\label{eq:protected}
    \big[\hat h^{\rm ee},\hat h^{\rm hh}\big](\varepsilon_0)=0 \quad\Leftrightarrow\quad[\bm J^{\rm e}_{\rm eff}\times\bm J^{\rm h}_{\rm eff}](\varepsilon_0)=\bm 0
\end{align}
then one can identify symmetry-protected bound states. Hence, a symmetry-protected crossing at $\varepsilon_0$ requires parallel effective fields and matching on-shell eigenvalues
\begin{align}
    [\mu_{\rm e} - \mu_{\rm h}] - [s|\bm J^{\rm e}_{\rm eff}| + t|\bm J^{\rm h}_{\rm eff}|]=0.
\end{align}
In case of a BCS substrate we have $\bm J_{\rm eff}^{\rm e}=\bm J_{\rm eff}^{\rm h}=\bm 0$ at the QPT which follows from Eq.~\eqref{eq:J_eff} and is depicted in Fig.~\ref{fig:effective_spinsplitting_BCS} which fulfils the symmetry condition in Eq.~\eqref{eq:protected}.

\subsection{Effective Hamiltonian close to the QPT} \label{sec:qpt}

\begin{figure}[ht]
	\centering
	\includegraphics[scale=1]{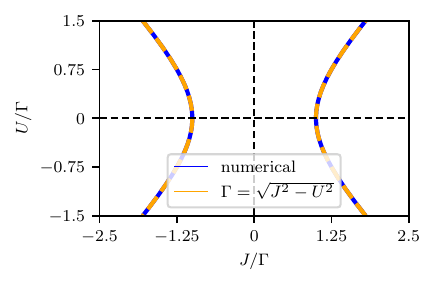}
	\caption{QPT condition as a function of the on-site potential $U$ and the impurity moment $J$ for a BCS substrate computed using the BCS limit in the numerical code and the analytical result.}
	\label{fig:qpt_bcs}
\end{figure}

Since an ISC generally breaks spin-rotation invariance for $B_x\neq0$, we introduce $\hat U_{\rm YSR}=\exp(-i\gamma \bm n\cdot\hat{\bm{\sigma}}/2)$ with $\bm n=\bm g \times\bm z/ |\bm g \times\bm z|$, where $\bm g$ denotes the normal state spin-vector in the electron block of the $4\times4$ Green's function and $\gamma$ is the rotation angle. We then introduce the effective single-particle Hamiltonian near the QPT at $\varepsilon=0$ and use it to construct the many-body Hamiltonian. For a single fermionic level, the Fock space reads $\mathcal H_{\rm mb}=\{\ket{0}, \ket{\uparrow}, \ket{\downarrow}, \ket{\uparrow\downarrow}\}$. After integrating out the bath (superconductor) dofs, the effective model is quadratic in the impurity operators $d^\dag_\sigma,d^{\phantom{\dag}}_\sigma$ and breaks particle number conservation through locally induced pairing but conserves parity. Hence, the many-body Hilbert space decouples into an even and an odd parity sector. With the convention
\begin{align}
	H_{\rm eff} = \sum_\sigma \varepsilon_\sigma d^{\phantom{\dag}}_\sigma d^\dag_\sigma + \Gamma_{\uparrow\downarrow} d_\uparrow d_{\downarrow} + \Gamma^*_{\uparrow\downarrow} d^\dag_\uparrow d^\dag_{\downarrow},
\end{align}
where $\Gamma_{\downarrow\uparrow}=-\Gamma_{\uparrow\downarrow}$ by hermiticity, the even subspace is described by
\begin{align}
	H_{\rm even} = \begin{bmatrix} \bra{0} & \bra{\uparrow\downarrow} \end{bmatrix} \begin{bmatrix} 0 & \Gamma_{\uparrow\downarrow} \\ \Gamma^*_{\uparrow\downarrow} & \varepsilon_\uparrow + \varepsilon_\downarrow \end{bmatrix}\begin{bmatrix} \ket{0} \\ \ket{\uparrow\downarrow} \end{bmatrix}
\end{align}
with eigenvalues
\begin{align}
	E^{\rm even}_\pm = \frac{\varepsilon_\uparrow + \varepsilon_\downarrow}{2} \pm \frac12\sqrt{[\varepsilon_\uparrow + \varepsilon_\downarrow]^2 + 4 |\Gamma_{\uparrow\downarrow}|^2}.
\end{align}
After rotating onto the dressed impurity spin quantization axis, the odd parity sector is described by
\begin{align}
	H_{\rm odd} = \begin{bmatrix} \bra{\downarrow} & \bra{\uparrow} \end{bmatrix} \begin{bmatrix} \varepsilon_\downarrow & 0 \\ 0 & \varepsilon_\uparrow \end{bmatrix}\begin{bmatrix} \ket{\downarrow} \\ \ket{\uparrow} \end{bmatrix}
\end{align}
with eigenvalues
\begin{align}
	E^{\rm odd}_\pm = \frac{\varepsilon_\uparrow + \varepsilon_\downarrow}{2} \pm \frac{\varepsilon_\uparrow - \varepsilon_\downarrow}{2}.
\end{align}
Since single-particle states reflect transitions between many-body states, a crossing of the many-body states enforces a zero-energy crossing in the single-particle spectrum. Using the lowest energy states of each parity sector, we define $E^{\rm YSR}_- \equiv E^{\rm odd}_- - E^{\rm even}_-$ and arrive at the general condition for a QPT in YSR systems
\begin{align}
	\label{eq:qptcond}
	|\Gamma_{\uparrow\downarrow}|^2 +\varepsilon_\uparrow\varepsilon_\downarrow = 0.
\end{align}
Note that without rotating onto the spin-quantization axis of the dressed impurity there are matrix elements $\varepsilon_{\sigma\bar\sigma}\neq0$ and spin is not a good description.
\paragraph{Example: BCS YSR states.} For a BCS superconductor~\cite{PhysRevB.103.155407}, one has $\hat g_{\rm YSR} = \hat g_{\uparrow\uparrow}\oplus \hat g_{\downarrow\downarrow}$ with
\begin{align}
	\hat g_{\sigma\sigma}(\varepsilon) = \frac{1}{\text{De}_\sigma}\Big[ &\big[\varepsilon\Gamma+(\varepsilon-J_\sigma)\sqrt{\Delta^2-\varepsilon^2} \big]\hat\tau_0 \notag\\
	\label{eq:BCS_YSR}
	& + \Gamma\Delta\hat\tau_x + U \sqrt{\Delta^2-\varepsilon^2}\hat\tau_z\Big]
\end{align}
in the basis $\bm d_{\rm imp}=[d_\uparrow, d_\downarrow^\dag,d_\downarrow,-d^\dag_\uparrow]$. The denominator $\text{De}_\sigma$ reads
\begin{align}
	\text{De}_\sigma = 2\Gamma \varepsilon[\varepsilon-J_\sigma] + [(\varepsilon-J_\sigma)^2-U^2-\Gamma^2]\sqrt{\Delta^2-\varepsilon^2}
\end{align}
which is the same as Eq.~\eqref{eq:bs_cond} for $U=0$ from before. The inverse Green's function follows straight forwardly from $[a \hat\tau_0 + b \hat\tau_x + c\hat\tau_z]^{-1}= [a \hat\tau_0 - b \hat\tau_x- c\hat\tau_z]/(a^2-b^2-c^2)$ and we find at the boundstate energy $\varepsilon=0$
\begin{subequations}
	\begin{align}
		\Gamma_{\uparrow\downarrow} &= -\frac{\Gamma}{J^2-U^2-\Gamma^2}, \\
		\varepsilon_\sigma &= -\frac{J_\sigma + U}{J^2-U^2-\Gamma^2}.
	\end{align}
\end{subequations}
Inserting into Eq.~\eqref{eq:qptcond} gives the relation $\Gamma = \sqrt{J^2-U^2}$ which we plot in Fig.~\ref{fig:qpt_bcs}. For $U=0$ it boils down to the well-known condition $\Gamma=J$ for the QPT from a singlet to a doublet ground state in BCS YSR systems. Note that for a BCS substrate, the condition is independent of the sign of $J$ and $U$ and that the QPT occurs for $J>\Gamma$.

\section{Computation of supercurrent and ABS spectrum}
\label{sec:current}
According to the definitions in the main text and with the BCS Green's function from Eq.~\eqref{eq:bcs_mybasis}, we have
\begin{align}
	\hat\Sigma_t = -\frac{\varepsilon\Gamma_t}{\sqrt{\Delta^2-\varepsilon^2}}\hat\tau_0\hat\sigma_0 - \frac{\Delta\Gamma_t e^{i\varphi\hat\tau_3\hat\sigma_0}}{\sqrt{\Delta^2-\varepsilon^2}} \hat\tau_2\hat\sigma_y.
\end{align}
The commutator in Eq.~\eqref{eq:supercurrent} thus boils down to
\begin{align}
	[\hat\Sigma_t, \hat G_{\rm imp}] = -\frac{\Delta\Gamma_t}{\sqrt{\Delta^2-\varepsilon^2}} \big[\hat X, \hat G_{\rm imp} \big],
\end{align}
where we have introduced
\begin{align}
	\hat X\equiv e^{i\varphi\hat\tau_3\hat\sigma_0}\hat\tau_2\hat\sigma_y = [\cos(\varphi)\hat\tau_2 + \sin(\varphi)\hat\tau_1]\hat\sigma_y.
\end{align}
For $\alpha,\gamma\in\{1,2,3\}$ and $\beta,\delta\in\{x,y,z\}$, it follows with
\begin{align}
	[\hat\tau_\alpha\hat\sigma_\beta, \hat\tau_\gamma\hat\sigma_\delta] = 2i [\varepsilon_{\alpha\gamma\eta}\delta_{\beta\delta} \hat\tau_\eta\hat\sigma_0 + \delta_{\alpha\gamma}\varepsilon_{\beta\delta\eta} \hat\tau_0\hat\sigma_\eta]
\end{align}
that only the coefficients multiplying $\hat\tau_1\hat\sigma_y, \hat\tau_2\hat\sigma_y$ in $\hat G_{\rm imp}$ contribute to $\tr\,(\hat\tau_3[\hat X,\hat G_{\rm imp}])$. With $G_{\alpha\beta}\equiv\tr\,(\hat\tau_\alpha\hat\sigma_\beta \hat G_{\rm imp})/4$, the supercurrent equivalently reads
\begin{align}
	\label{eq:bcs_current}
	I(\varphi) = \frac{8e\Delta\Gamma_t}{\hbar} \int_{\mathbb R}d\varepsilon \Re\,\left\{ i f(\varepsilon)\frac{G^a_{2y}\sin(\varphi) - G^a_{1y}\cos(\varphi)}{\sqrt{\Delta^2-[\varepsilon + i\eta]^2}} \right\}.
\end{align}
\begin{figure*}[ht]
    \centering
    \includegraphics[scale=1]{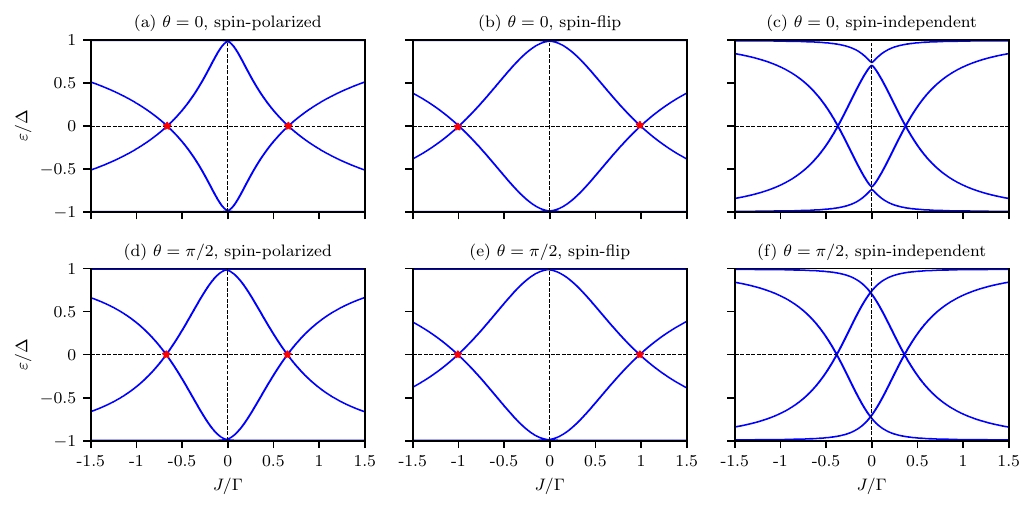}
    \caption{Bound state energies for $\{\beta,B_x,U_0, \pi N_0|v_1|^2\}=\{10,1,0,50\}\Delta_0$, $|v_1|^2/|v_2|^2=5$, $\theta=0,\pi/2$ in (a)--(c), (d)--(f), respectively, and the indicated hopping cases. The gap is calculated self-consistently for $T=10^{-3}T_c$. Red stars indicate symmetry-protected state crossings as outlined in Sec.~\ref{sec:bs_symmetry}.}
    \label{fig:ysrstates2}
\end{figure*}
\section{\texorpdfstring{Bound state spectrum for experimentally relevant values of $B_x$}{}}
In Fig.~\ref{fig:ysrstates2}, we present the YSR bound state spectrum for experimentally accessible values of magnetic field strengths $B_x/\beta=0.1$ since the magnitude of ISOC $\beta$ is generally large compared to experimentally accessible magnetic fields~\cite{PhysRevB.106.184514,doi:10.1126/science.aab2277}. As in Fig.~\ref{fig:ysrstates}, we include plots for the impurity realizations $\theta=0,\pi/2$ and the three cases of hopping from Sec.~\ref{sec:hybridization}.


%

\end{document}